\documentclass[preprint,5p,times]{elsarticle}

\usepackage{subcaption}
\usepackage{caption} 

\usepackage{graphicx}

\usepackage{tikz}
\usepackage{pgfplots}
\usepackage{booktabs}

\usepackage{multirow}

\usepackage{placeins}

\usepackage{amsmath}
\usepackage{amsthm}
\usepackage{amssymb}
\usepackage{amsfonts}
\usepackage{mathtools}
\usepackage{microtype}

\usepackage{tikz}
\usetikzlibrary{trees}
\pgfplotsset{compat=1.18}

\biboptions{numbers,sort&compress}

\newcommand{\convergenceorder}{\ensuremath{n}}
\newcommand{\stepsize}{{\ensuremath{h}}}

\newcommand{\Hamiltonian}{\ensuremath{\mathcal{H}}}
\newcommand{\Potential}{\ensuremath{\mathcal{V}}}
\newcommand{\Action}{\ensuremath{\mathcal{S}}}
\newcommand{\Kinetic}{\ensuremath{\mathcal{T}}}

\newcommand{\Potentialvf}{\ensuremath{\hat{\Potential}}}
\newcommand{\Actionvf}{\ensuremath{\hat{\Action}}}
\newcommand{\Kineticvf}{\ensuremath{\hat{\Kinetic}}}
\newcommand{\ForceGradientvf}{\ensuremath{\hat{\mathcal{C}}}}

\newcommand{\EffErr}[1]{\ensuremath{\mathrm{Eff}_{\mathtt{err}}^{(#1)}}}
\newcommand{\EffStab}{\ensuremath{\mathrm{Eff}_{\mathtt{stab}}}}

\newcommand{\p}{\ensuremath{\mathbf{p}}}
\newcommand{\q}{\ensuremath{\mathbf{q}}}
\newcommand{\massmatrix}{\ensuremath{\mathbf{M}}}
\newcommand{\stabilitymatrix}{\ensuremath{\mathbf{K}}}
\newcommand{\0}{\ensuremath{\mathbf{0}}}
\DeclareMathOperator{\e}{\mathbf{e}}

\DeclareMathOperator{\trace}{tr}

\newcommand{\LieGroup}{\ensuremath{\mathcal{G}}}

\newcommand{\SUthree}{\ensuremath{\mathrm{SU}(3)}}
\newcommand{\suthree}{\ensuremath{\mathfrak{su}(3)}}
\newcommand{\Generator}{\ensuremath{\boldsymbol{T}}}
\newcommand{\U}{\ensuremath{\boldsymbol{U}}}
\newcommand{\MatrixP}{\ensuremath{\boldsymbol{P}}}
\newcommand{\diffoperator}{\ensuremath{\boldsymbol{e}}}

\newtheorem{example}{Example}
\newtheorem{definition}{Definition}
\newtheorem{remark}{Remark}
\newtheorem{proposition}{Proposition}

\usepackage[colorlinks,breaklinks]{hyperref}

\journal{Computer Physics Communications}

\begin{document}

\begin{frontmatter}
\title{Numerical stability of force-gradient integrators and their Hessian-free variants in lattice QCD simulations}

\author[1]{Kevin Schäfers\corref{cor1}}
\ead{schaefers@math.uni-wuppertal.de}
\author[3,2]{Jacob Finkenrath}
\ead{finkenrath@uni-wuppertal.de}
\author[1]{Michael Günther}
\ead{guenther@math.uni-wuppertal.de}
\author[2]{Francesco Knechtli}
\ead{knechtli@physik.uni-wuppertal.de}

\cortext[cor1]{Corresponding author.}

\affiliation[1]{organization={Institute of Mathematical Modelling, Analysis and Computational Mathematics (IMACM), Chair of Applied and Computational Mathematics, Bergische Universität Wuppertal},
            addressline={Gaußstraße 20}, 
            city={Wuppertal},
            postcode={42119}, 
            country={Germany}}
\affiliation[2]{organization={Department of Physics, Bergische Universität Wuppertal},
            addressline={Gaußstraße 20}, 
            city={Wuppertal},
            postcode={42119}, 
            country={Germany}}
\affiliation[3]{organization={Department of Theoretical Physics},
            addressline={European Organization for Nuclear
Research, CERN}, 
            city={Geneve},
            postcode={1211}, 
            country={Switzerland}}

\begin{abstract}
A comprehensive linear stability analysis of force-gradient integrators and their Hessian-free variants is carried out by investigating the harmonic oscillator as a test equation.
The analysis reveals that the linear stability of conventional force-gradient integrators and their Hessian-free counterparts coincides.
By performing detailed linear stability investigations for the entire family of self-adjoint integrators with up to eleven exponentials per time step, we detect promising integrator variants that are providing a good trade-off between accuracy and numerical stability.
Special attention is given to the application of these promising integrator variants within the Hamiltonian Monte Carlo algorithm, particularly in the context of interacting field theories.
Simulations for the two-dimensional Schwinger model are conducted to demonstrate that there are no significant differences in the stability domain of a force-gradient integrator and its Hessian-free counterpart.
Lattice QCD simulations with two heavy Wilson fermions emphasize that Hessian-free force-gradient integrators with a larger stability threshold allow for a more efficient computational process compared to conventional splitting methods. 
Furthermore, detailed investigations of the stability threshold are performed by investigating $N_f = 2$ twisted-mass fermions and nested integrators, highlighting the reliability of the linear stability threshold for lattice QCD simulations.
\end{abstract}

\end{frontmatter}


\section{Introduction}
The Hamiltonian Monte Carlo (HMC) algorithm \cite{duane1987hybrid} is a frequently employed algorithm for lattice QCD simulations.
In the molecular dynamics (MD) step of the HMC algorithm, trial configurations are generated by integrating the Hamiltonian equations of motion from an initial configuration for a fictitious time $\tau$.
In a subsequent Metropolis step \cite{metropolis1953}, configurations are accepted or rejected by investigating the energy change $\Delta \Hamiltonian$ along a trajectory.

The selection of the numerical integration scheme is crucial for an efficient and structure-preserving integration of the equations of motion.
Firstly, the numerical integration scheme must be volume-preserving and time-reversible to satisfy the detailed balance condition \cite{duane1987hybrid}, ensuring that the Markov process converges to the correct equilibrium.
Secondly, the links are elements of the special unitary group $\SUthree$, and therefore we additionally require that the integrator provides numerical approximations situated on the Lie group manifold.  

\subsection{Decomposition algorithms}
For separable Hamiltonian systems \mbox{$\Hamiltonian(\q,\p) = \Kinetic(\p) + \Potential(\q)$}, decomposition algorithms \cite{mclachlan2002splitting,omelyan2003symplectic} facilitate the derivation of explicit numerical integration schemes that adhere to the aforementioned properties. In addition to conventional splitting methods \cite{mclachlan2002splitting}, force-gradient integrators \cite{omelyan2003symplectic} enhance the computational process by incorporating a symplectic corrector \cite{wisdom1996symplectic} into the momentum updates. 
The correction necessitates the inclusion of the \emph{force-gradient term}, which involves the Hessian of the potential. 
Hessian-free force-gradient integrators \cite{schäfers2024hessianfree} utilize an approximation of the force-gradient updates, effectively replacing the force-gradient term in the force-gradient step with a second evaluation of the force $-\Potential_{\q}(\q)$. 
The efficiency of the Hessian-free variants has been demonstrated in various large-scale lattice QCD simulations, see e.g.~\cite{Jung2017xef,Finkenrath2022eon,Finkenrath2023sjg,schäfers2024hessianfree}.

\subsection{Tuning integrators}
When tuning a numerical integration scheme for lattice QCD simulations, we aim at minimizing the computational cost to achieve the optimal acceptance rate of $\langle P_{\mathrm{acc}}\rangle_{\mathrm{opt}} = \exp(-1/\convergenceorder)$ with $\convergenceorder$ denoting the convergence order of the numerical integration scheme \cite{TAKAISHI20006}. 
For normally distributed $\Delta \Hamiltonian$, the acceptance rate corresponds to the variance $\sigma^2(\Delta \Hamiltonian)$ via 
$$\langle P_{\mathrm{acc}} \rangle = \mathrm{erfc}\left( \sqrt{\sigma^2(\Delta \Hamiltonian)/8} \right),$$
where $\mathrm{erfc}$ denotes the complementary error function \cite{KNECHTLI20033}. 
Consequently, one aims at maximizing the step size $\stepsize$ while satisfying 
\begin{equation}\label{eq:accuracy_criterion} 
    \sigma^2(\Delta \Hamiltonian) \le 8 \cdot \mathrm{erfc}^{-1}(\langle P_{\mathrm{acc}} \rangle_{\mathrm{opt}})^2,
\end{equation}
i.e., we are primarily concerned with the region 
$$10^{-2} \le \sigma^2(\Delta \Hamiltonian) \le 1,$$
which corresponds to acceptance probabilities ranging from from 61.7\% to 96\%.

\subsection{Numerical stability}
During the tuning process, the numerical stability of the integrator assumes paramount importance. Decomposition algorithms applied to separable Hamiltonian systems are only conditionally stable, necessitating the selection of a sufficiently small step size $\stepsize$.
The primary challenge lies in identifying whether the accuracy or the numerical stability of the numerical integration scheme becomes the limiting factor. 
In such instances, a superior integrator can be obtained by selecting an algorithm that is more accurate or more stable, respectively.

While the accuracy of decomposition algorithms has been extensively studied \cite{omelyan2003symplectic,schäfers2024hessianfree}, the numerical stability of decomposition algorithms has only been investigated for conventional splitting methods \cite{BlanesStability}.
Since (Hessian-free) force-gradient integrators turn out to be more efficient decomposition algorithms for increasing lattice volumes, an investigation of the numerical stability of (Hessian-free) force-gradient integrators is of particular interest.
It is commonly argued \cite{EDWARDS1997375,instabilityMD} that the instability in lattice QCD can be likened to that of a collection of oscillator modes, with the onset of the instability caused by the mode of highest frequency, $\omega_{\max}$.
Based on this hypothesis, the linear stability analysis of decomposition algorithms when applied to the harmonic oscillator is sufficient to study the numerical stability for lattice QCD simulations. 

The present paper contributes to providing a linear stability analysis of force-gradient integrators and their Hessian-free variants. This analysis enables the identification of accurate decomposition algorithms with a large stability domain within the family of self-adjoint algorithms.

\subsection{Outline}
The paper is organized as follows.
Section \ref{sec:integrator_survey} provides a brief survey of splitting methods and (Hessian-free) force-gradient integrators. We moreover show that force-gradient integrators and their Hessian-free variants are equivalent when applied to linear systems. Consequently, the linear stability of both frameworks is identical. 
Subsequently, Section \ref{sec:linear_stability_analysis} will undertake a linear stability analysis of force-gradient integrators by tailoring the existing linear stability analysis for conventional splitting methods \cite{BlanesStability} to the force-gradient approach.
Section \ref{sec:detection} will address the optimization of the algorithms with respect to stability and accuracy. It will be demonstrated that already small alterations to the integrator coefficients will result in a significant reduction in the efficiency measures employed in \cite{omelyan2003symplectic,schäfers2024hessianfree}. 
Without making any further assumptions on the potential $\Potential(\q)$, the most promising integrators are identified by computing the stability thresholds of all variants listed in \cite{omelyan2003symplectic,schäfers2024hessianfree} that are minimizing the norm of the leading error coefficients. 
As already mentioned, the linear stability of force-gradient integrators and their Hessian-free variants is identical. 
In contrast, for nonlinear systems such as those encountered in lattice QCD simulations, the two frameworks differ, potentially leading to differences in numerical stability. 
Consequently, the numerical tests presented in Section~\ref{sec:numerical_tests} start with an examination of potential discrepancies in the numerical stability of force-gradient integrators and their Hessian-free counterparts. This investigation is conducted within the two-dimensional Schwinger model, where an analytical expression of the force-gradient term is available~\cite{shcherbakov2017adapted}. These findings substantiate the absence of significant differences in the numerical stability.
Afterwards, we will consider an ensemble with two heavy Wilson fermions \cite{Knechtli2022,schäfers2024hessianfree} and demonstrate that the performance of the integrators is consistent with the stability analysis presented in this paper. 
Furthermore, numerical simulations will be conducted with $N_f = 2$ twisted-mass fermions,
providing numerical evidence that the linear stability analysis provides a reliable stability criterion for lattice QCD simulations.

\section{A brief survey of decomposition algorithms}\label{sec:integrator_survey}
In this section, we will briefly introduce splitting methods and (Hessian-free) force-gradient integrators for separable Hamiltonian systems of the form 
\begin{equation}\label{eq:Hamiltonian_system}
    \Hamiltonian(\q,\p) = \Kinetic(\p) + \Potential(\q) = \frac{1}{2} \p^\top \massmatrix \p + \Potential(\q) 
\end{equation}
with symmetric and positive definite matrix $\massmatrix \in \mathbb{R}^{d \times d}$. 
Since we will apply the integrators to the harmonic oscillator to perform the linear stability analysis, we consider the canonical case with phase space $\mathbb{R}^{2d}$ here. 
The formulations of the integrators when the phase space is the cotangent bundle $T^* \LieGroup$ over a configuration space manifold $\LieGroup$ that is a semi-simple matrix Lie group can be found in~\ref{app:LieGroup}.
For a more general formulation, we refer to \cite{schäfers2024hessianfree,kennedy2013shadow}.

\subsection{Hamiltonian equations of motion}
In the canonical case, the vector fields of the kinetic and potential energy zero-forms are given by 
\begin{align}\label{eq:kinetic_and_potential_vector_fields}
    \Kineticvf &= \sum_{i,j=1}^d m_{i,j} p_j \frac{\partial}{\partial q_i},   
    &
    \Potentialvf &= -\sum_{i=1}^d \Potential_{q_i}(\q) \frac{\partial}{\partial p_i}.
\end{align}
Consequently, the equations of motion read 
\begin{align*}
    \dot{\q} &= \Kineticvf \q = \massmatrix \p, 
    &
    \dot{\p} &= \Potentialvf \p = -\Potential_{\q}(\q).
\end{align*} 
The natural splitting $\Hamiltonian = \Kinetic + \Potential$ yields two easily solvable Hamiltonian systems, namely
\begin{align}\label{eq:subsystems}
    \begin{pmatrix}
        \dot{\q} \\ \dot{\p}
    \end{pmatrix} &= \begin{pmatrix}
        \massmatrix \p \\ \0
    \end{pmatrix}, 
    & 
    \begin{pmatrix}
        \dot{\q} \\ \dot{\p}
    \end{pmatrix} &= \begin{pmatrix}
        \0 \\ -\Potential_{\q}(\q)
    \end{pmatrix}. 
\end{align}
The exact flows of these two subsystems are
\begin{align}
    \e^{t \Kineticvf}\begin{pmatrix}
        \q_0 \\ \p_0
    \end{pmatrix} &= \begin{pmatrix}
        \q_0 + t\massmatrix\p_0 \\
        \p_0
    \end{pmatrix}, \label{eq:position_update} \\ 
    \e^{t \Potentialvf}\begin{pmatrix}
        \q_0 \\ \p_0
    \end{pmatrix} &= \begin{pmatrix}
        \q_0 \\
        \p_0 - t\Potential_{\q}(\q_0)
    \end{pmatrix}. \label{eq:momentum_update}
\end{align}
In this context, we refer to \eqref{eq:position_update} as a \emph{position update} and to \eqref{eq:momentum_update} as a \emph{momentum update}.

\subsection{Splitting methods}
In light of the splitting $\Hamiltonian = \Kinetic + \Potential$, splitting methods \cite{mclachlan2002splitting} compute a numerical approximation of the formal solution $\e^{\stepsize (\Kineticvf + \Potentialvf)}$ through a composition of the position updates \eqref{eq:position_update} and momentum updates \eqref{eq:momentum_update}. 
A $s$-stage splitting method is defined by arbitrary coefficients $a_1,b_1,\ldots,a_s,b_s \in \mathbb{R}$ and can be expressed as
\begin{equation}\label{eq:splitting_method}
    \Phi_\stepsize =
    \e^{a_1 \stepsize \Kineticvf}
    \e^{b_1 \stepsize \Potentialvf} \cdots
    \e^{a_{s-1} \stepsize \Kineticvf}
    \e^{b_{s-1} \stepsize \Potentialvf} \e^{a_s \stepsize \Kineticvf} \e^{b_s \stepsize \Potentialvf}.
\end{equation}
As a composition of symplectic maps, the resulting integrator is symplectic and thus volume-preserving.
To obtain a time-reversible integrator, it is necessary to consider self-adjoint splitting methods, i.e., splitting methods \eqref{eq:splitting_method} whose coefficients satisfy for all $\ell=1,\ldots,s$ either 
\begin{subequations}\label{eq:symmetry_relations}
    \begin{align}
        a_1 &= 0,\ a_{\ell+1} = a_{s-\ell+1}, \ \text{and } b_\ell = s_{s-\ell+1}, \label{eq:velocity-version} 
    \intertext{resulting in a \emph{velocity version}, or}
        b_s &= 0,\ b_\ell = b_{s-\ell},\ \text{and }a_\ell = a_{s-\ell+1}, \label{eq:position-version} 
    \end{align}
\end{subequations}
resulting in a \emph{position version}.
If the integrator coefficients satisfy
\begin{equation}\label{eq:order1-condition}
    \sum_{\ell=1}^s a_\ell = \sum_{\ell=1}^s b_\ell = 1,
\end{equation}
the splitting method is consistent and, due to the inherent symmetry, of convergence order $\convergenceorder \ge 2$. 
The order conditions for splitting methods can be determined by applying the Baker--Campbell--Hausdorff (BCH) formula \cite{mclachlan2002splitting}. 
A splitting method \eqref{eq:splitting_method} is of convergence order $\convergenceorder$ if 
$$ \Phi_\stepsize = \e^{\stepsize(\Kineticvf + \Potentialvf)} +\, \mathcal{O}(\stepsize^{\convergenceorder+1}).$$
For further details on the order theory and the construction of higher-order splitting methods, we refer to \cite{mclachlan2002splitting,Blanes_Casas_Murua_2024} and the references therein.

\subsection{Force-gradient integrators}
\begin{sloppypar}
For Hamiltonian systems of the form \eqref{eq:Hamiltonian_system}, splitting methods \eqref{eq:splitting_method} can be enhanced by incorporating the commutator \mbox{$\ForceGradientvf \coloneqq [\Potentialvf,[\Kineticvf,\Potentialvf]]$} into the computational process. 
Given the vector fields \eqref{eq:kinetic_and_potential_vector_fields}, the commutator solely depends on $\q$ and reads  
\end{sloppypar}
\begin{equation}\label{eq:force-gradient_term}
    \ForceGradientvf = 2 \sum_{i,j,k = 1}^d \Potential_{q_k}(\q) m_{j,k} \Potential_{q_i q_j}(\q) \frac{\partial}{\partial p_i}.
\end{equation}
Since the term involves the gradient of the force, it is frequently referred to as the \emph{force-gradient term}.
The incorporation of force-gradient evaluations into the momentum updates \eqref{eq:momentum_update} results in force-gradient steps 
\begin{align}\label{eq:force-gradient_step}
\begin{split}
&\e^{b_\ell \stepsize \Potentialvf + c_\ell \stepsize^3 \ForceGradientvf}\begin{pmatrix}
    \q_0 \\ \p_0
\end{pmatrix} \\
&\qquad\coloneqq \begin{pmatrix}
    \q_0 \\ \p_0 - b_\ell \stepsize \Potential_{\q}(\q_0) + 2c_\ell \stepsize^3 \Potential_{\q\q}(\q_0) \massmatrix \Potential_{\q}(\q_0)
\end{pmatrix}. 
\end{split}
\end{align}
A force-gradient integrator \cite{kennedy2009force,omelyan2003symplectic} with $s$ stages reads
\begin{equation}\label{eq:force-gradient-integrator}
    \Phi_\stepsize = \e^{a_1 \stepsize \Kineticvf} \e^{b_1 \stepsize \Potentialvf + c_1 \stepsize^3 \ForceGradientvf} \cdots \e^{a_s \stepsize \Kineticvf} \e^{b_s \stepsize \Potentialvf + c_s \stepsize^3 \ForceGradientvf}. 
\end{equation}
A force-gradient step \eqref{eq:force-gradient_step} involving the force-gradient term $\ForceGradientvf$ may be regarded as a conventional momentum update \eqref{eq:momentum_update}, applied to a Hamiltonian with modified potential 
$$\tilde{\mathcal{V}}(\q) = \Potential(\q) - \frac{c_\ell \stepsize^2}{b_\ell} \mathcal{C}(\q),$$
where $\mathcal{C}(\q)$ is solved by 
$\frac{\partial}{\partial q_i} \mathcal{C}(\q) = 2 \sum_{j,k=1}^d \Potential_{q_k}(\q) m_{j,k} \Potential_{q_i q_j}(\q)$. 
Moreover, the modified potential is again Hamiltonian, i.e., the exact flow is again a symplectic map. Thus, force-gradient integrators are symplectic and, as a consequence, volume-preserving. 
Self-adjoint force-gradient integrators, that is, those integrators satisfying the symmetry relations \eqref{eq:symmetry_relations} and where the coefficients $c_\ell$ moreover satisfy the same symmetry relations as the coefficients $b_\ell$, are time-reversible.
For a comprehensive analysis of the order theory of force-gradient integrators, including a detailed classification of self-adjoint integrators with $s \leq 6$ stages, we refer to \cite{omelyan2003symplectic}.

\subsection{Hessian-free force-gradient integrators}
If computing the force-gradient update \eqref{eq:force-gradient_step} is significantly more expensive than computing a momentum update \eqref{eq:momentum_update}, or if an expression for \eqref{eq:force-gradient_term} is not available (either analytically or in a software implementation), then \emph{Hessian-free force-gradient integrators} \cite{schäfers2024hessianfree} represent a promising alternative. 
In this approach, the force-gradient step \eqref{eq:force-gradient_step} is approximated via 
\begin{align}\label{eq:modified_momentum_update}
    \begin{split}
    \e^{b_\ell \stepsize \hat{\mathcal{D}}(\stepsize,b_\ell,c_\ell)}\!\begin{pmatrix}
        \q_0\\\p_0
    \end{pmatrix} &\coloneqq \begin{pmatrix}
        \q_0 \\ \p_0 - b_\ell \stepsize \Potential_{\q}\left(\q_0 - \frac{2c_\ell \stepsize^2}{b_\ell} \massmatrix \Potential_{\q}(\q_0) \right)
    \end{pmatrix} \\
    &= \e^{b_\ell \stepsize \Potentialvf + c_\ell \stepsize^3 \ForceGradientvf}\begin{pmatrix}
            \q_0 \\ \p_0
    \end{pmatrix} + \mathcal{O}(\stepsize^5),
    \end{split}
\end{align}
effectively replacing the Hessian of the potential by an additional force evaluation. 
Consequently, a $s$-stage Hessian-free force-gradient integrator is given by
\begin{equation}\label{eq:Hessian-free_FGI}
    \Phi_h = \e^{a_1 \stepsize \Kineticvf} \e^{b_1 \stepsize \hat{\mathcal{D}}(\stepsize,b_1,c_1)} \cdots \e^{a_s \stepsize \Kineticvf} \e^{b_s \stepsize \hat{\mathcal{D}}(\stepsize,b_s,c_s)}.
\end{equation}
The approximation of the force-gradient step does not affect the time-reversibility of the method, provided that the composition is self-adjoint. 
As a composition of shears, Hessian-free force-gradient integrators are also volume-preserving. The approximated force-gradient step \eqref{eq:modified_momentum_update}, however, is no longer a symplectic map, except in the special cases where either $c_\ell = 0$ or the Hessian of the potential commutes with itself when evaluated at (in general, different) $\q$.
A consequence of symplecticity is the preservation of a nearby shadow Hamiltonian \cite{kennedy2013shadow}. In contrast to force-gradient integrators, the Hessian-free variants do not preserve a shadow Hamiltonian. In general, without any additional assumptions on the potential, the numerical energy will thus exhibit a drift of size $\mathcal{O}(\tau \stepsize^{\max\{4,\convergenceorder\}})$. The lack of long-time energy conservation of the Hessian-free variants may prove disadvantageous, particularly for exponentially long-time simulations. In lattice QCD, however, trajectory lengths of $\tau \approx 2$ are typically considered. Therefore, the theoretical energy drift has no impact on the acceptance probability of the HMC algorithm.
A Taylor series expansion of the approximation \eqref{eq:modified_momentum_update} reveals that the additional error terms involve derivatives of the potential $\Potential$ of order three and higher. It follows that for quadratic Hamiltonian functions, i.e., Hamiltonian functions \eqref{eq:Hamiltonian_system} with potential energy $\Potential(\q) = \frac{1}{2} \q^\top \mathbf{N} \q$, the approximated force-gradient update \eqref{eq:modified_momentum_update} is identical to the force-gradient step \eqref{eq:force-gradient_step}. 
As a consequence, the two frameworks are equivalent when applied to linear ODE systems like the harmonic oscillator, i.e., their linear stability is identical.
\remark{
To identify splitting methods, force-gradient integrators, and Hessian-free force-gradient integrators, we will use the following abbreviations. $A$ denotes a position update \eqref{eq:position_update}, $B$ a momentum update \eqref{eq:momentum_update} with $c_{\ell} = 0$, $C$ a conventional force-gradient step \eqref{eq:force-gradient_step} with $c_{\ell} \neq 0$, and $D$ an approximated force-gradient step \eqref{eq:modified_momentum_update} with $c_{\ell} \neq 0$.
}\normalfont\medskip

Without loss of generality, we will carry out a linear stability analysis for force-gradient integrators~\eqref{eq:force-gradient-integrator}.

\section{Linear stability of force-gradient integrators}
\label{sec:linear_stability_analysis}
As the exact flows of the subsystems \eqref{eq:subsystems} are not bounded, the decomposition algorithms presented in Section \ref{sec:integrator_survey} are only conditionally stable. 
To perform a numerical stability analysis, one selects a simple test problem with bounded solutions whose flow lies in the same subgroup of the group of diffeomorphisms. 
In case of Hamiltonian systems, this is typically the harmonic oscillator
\begin{subequations}\label{eq:harmonic_oscillator}
\begin{equation}\label{eq:harmonic_oscillator_second-order}
    \ddot{y} + \omega^2 y = 0, \quad \omega > 0.
\end{equation}
\begin{sloppypar}
For the linear stability analysis, we consider the splitting \mbox{$(q,p) = (\omega y,\dot{y})$}, which gives rise to the first-order system 
\end{sloppypar}
\begin{equation}\label{eq:harmonic_oscillator_first-order}
    \begin{pmatrix}
        \dot{q} \\ \dot{p}
    \end{pmatrix} = \Bigg[ \begin{pmatrix}
        0 & \omega \\ 0 & 0 
    \end{pmatrix} + \begin{pmatrix}
        0 & 0 \\ -\omega & 0 
    \end{pmatrix} \Bigg] \begin{pmatrix}
        q \\ p
    \end{pmatrix} = \begin{pmatrix}
        \omega p \\ 0
    \end{pmatrix} + \begin{pmatrix}
        0 \\ -\omega q
    \end{pmatrix}.
\end{equation}
\end{subequations}
The force-gradient term \eqref{eq:force-gradient_term} for the harmonic oscillator \eqref{eq:harmonic_oscillator_first-order} reads
$\ForceGradientvf = 2 \omega^3 q \frac{\partial}{\partial p}$.
A decomposition algorithm will typically become unstable for $\lvert z \rvert > z_*$, where $z \coloneqq \omega \stepsize$ and the parameter $z_*$ represents the \emph{stability threshold} of the integrator. 
The objective is to determine the parameter $z_*$ for a given force-gradient integrator by adapting the theory for splitting methods as presented in \cite{BlanesStability}.
Starting from the initial values $(q(0),p(0)) = (q_0,p_0)$, the exact solution to the harmonic oscillator \eqref{eq:harmonic_oscillator_first-order} at time point $\stepsize > 0$ is given by 
\begin{equation*}
    \begin{pmatrix}
        q(\stepsize) \\ p(\stepsize) 
    \end{pmatrix} = \underbrace{\begin{pmatrix}
        \cos(z) & \sin(z) \\ -\sin(z) & \cos(z)
    \end{pmatrix}}_{= \mathbf{O}(z)} \begin{pmatrix}
        q_0 \\ p_0
    \end{pmatrix}, \quad z \coloneqq \omega \stepsize.
\end{equation*}
Applying a force-gradient integrator \eqref{eq:force-gradient-integrator} to the harmonic oscillator \eqref{eq:harmonic_oscillator_first-order} yields an approximation 
$$
\begin{pmatrix}
    q_1 \\ p_1
\end{pmatrix} = \stabilitymatrix(z) \begin{pmatrix}
    q_0 \\ p_0
\end{pmatrix}
$$
with \emph{stability matrix}
\begin{align}\label{eq:FGI_applied_to_osc}
    \begin{split}
        \stabilitymatrix(z) &= \prod_{\ell = 1}^s \begin{pmatrix}
            1 & 0 \\ -b_{s+1-\ell} z + 2c_{s+1-\ell} z^3 & 1 
        \end{pmatrix}  \begin{pmatrix}
            1 & a_{s+1-\ell} z \\ 0 & 1
        \end{pmatrix},
    \end{split}
\end{align}
which is an approximation to $\mathbf{O}(z)$. 
Additionally, we define the \emph{stability polynomial}
\begin{equation}\label{eq:stability-polynomial}
    \mathfrak{p}(z) \coloneqq \frac{1}{2} \trace(\stabilitymatrix(z)).
\end{equation}
In case of self-adjoint force-gradient integrators, the stability matrix \eqref{eq:FGI_applied_to_osc} takes the form 
\begin{subequations}
\begin{equation}\label{eq:stability_matrix}
    \stabilitymatrix(z) = \begin{pmatrix}
        \mathfrak{p}(z) & K_{1,2}(z) \\ K_{2,1}(z) & \mathfrak{p}(z)
    \end{pmatrix},
\end{equation}
with even stability polynomial 
\begin{equation}\label{eq:stability-polynomial_self-adjoint}
    \mathfrak{p}(z) = 1 + \sum\limits_{i=1}^{2(s-1)} \bar{p}_i z^{2i},
\end{equation}
and odd polynomials 
\begin{equation}\label{eq:odd-polynomials_in_K}
    K_{1,2}(z) = \sum_{i=1}^{2s-1} k_i z^{2i-1}, \quad K_{2,1}(z) = \sum_{i=1}^{2s} \bar{k}_{i} z^{2i-1},
\end{equation}
\end{subequations}
where $\bar{p}_i,k_i,\bar{k}_i$ are homogeneous polynomials in the coefficients $a_\ell,b_\ell,c_\ell$. 
In particular, it holds 
$$ k_{2s-2} = k_{2s-1} = 0 $$
for a velocity version \eqref{eq:velocity-version}, and
$$ \bar{k}_{2s-1} = \bar{k}_{2s} = 0 $$
for a position version \eqref{eq:position-version} of the force-gradient integrator \eqref{eq:force-gradient-integrator}. 
Furthermore, for consistent schemes satisfying \eqref{eq:order1-condition}, it holds 
$$ \bar{p}_1 = -\tfrac{1}{2}, \quad k_1 = 1, \quad \bar{k}_1 = -1, $$
resulting in $\mathfrak{p}(z) = 1 - z^2/2 + \mathcal{O}(z^4)$.

\example{
Consider the force-gradient integrator BACAB
\begin{subequations}\label{eq:BAXAB}
\begin{equation}\label{eq:BACAB}
    \Phi_\stepsize = \e^{\tfrac{\stepsize}{6} \Potentialvf} \e^{\tfrac{\stepsize}{2} \Kineticvf} \e^{\tfrac{2\stepsize}{3} \Potentialvf + \tfrac{\stepsize^3}{72} \ForceGradientvf} \e^{\tfrac{\stepsize}{2} \Kineticvf} \e^{\tfrac{\stepsize}{6} \Potentialvf}, 
\end{equation}
and its Hessian-free variant BADAB
\begin{equation}\label{eq:BADAB}
    \Phi_\stepsize = \e^{\tfrac{\stepsize}{6} \Potentialvf} \e^{\tfrac{\stepsize}{2} \Kineticvf} \e^{\tfrac{2\stepsize}{3} \hat{\mathcal{D}}\left(\stepsize,\tfrac{2}{3},\tfrac{1}{72}\right)} \e^{\tfrac{\stepsize}{2} \Kineticvf} \e^{\tfrac{\stepsize}{6} \Potentialvf}. 
\end{equation}
\end{subequations}
The stability matrix of these two integrators is given by 
$$ \stabilitymatrix(z) = \begin{pmatrix}
    1 - \tfrac{z^2}{2} + \tfrac{z^4}{24} - \tfrac{z^6}{864} & z - \tfrac{z^3}{6} + \tfrac{z^5}{144} \\
    -z + \tfrac{z^3}{6} - \tfrac{z^5}{108} + \tfrac{z^7}{5184} &  1 - \tfrac{z^2}{2} + \tfrac{z^4}{24} - \tfrac{z^6}{864}
\end{pmatrix}, $$
with stability polynomial $\mathfrak{p}(z) = 1 - \tfrac{z^2}{2} + \tfrac{z^4}{24} - \tfrac{z^6}{864}$.
}\normalfont\medskip
The stability analysis presented in \cite{BlanesStability} can be applied to force-gradient integrators in a straightforward manner. 
We provide a brief summary of the fundamental results, tailored to the framework of force-gradient integrators. 
A force-gradient integrator \eqref{eq:force-gradient-integrator} is considered as stable when applied to the harmonic oscillator \eqref{eq:harmonic_oscillator_first-order} if $\stabilitymatrix(z)^i$ is bounded independently of the number of time steps $i \ge 1$. 
\definition[Stability interval]{We denote by $z_*$ the largest non-negative real number such that $\stabilitymatrix(z)$ is stable for all $z \in (-z_*,z_*)$. We refer to $z_*$ as the stability threshold and call $(-z_*,z_*)$ the stability interval of $\stabilitymatrix(z)$.}\normalfont\medskip 
We moreover introduce the following characterization of stability.
\proposition{
Let $\stabilitymatrix(z)$ denote the stability matrix of a self-adjoint force-gradient integrator \eqref{eq:force-gradient-integrator} and $\mathfrak{p}(z)$ its stability polynomial. Then, the following statements are equivalent:
\begin{enumerate}
    \item[\rm (i)] The matrix $\stabilitymatrix(z)$ is stable.
    \item[\rm (ii)] The matrix $\stabilitymatrix(z)$ is diagonalizable with eigenvalues of modulus one.
    \item[\rm (iii)] It holds $\lvert \mathfrak{p}(z) \rvert \le 1$, and $\stabilitymatrix(z)$ is similar to the matrix
    $$ \mathbf{S}(z) = \begin{pmatrix}
        \cos{\Psi(z)}  & \sin{\Psi(z)} \\
        -\sin{\Psi(z)} & \cos{\Psi(z)}
    \end{pmatrix}, \, \text{where } \Psi(z) = \arccos{\mathfrak{p}(z)}.$$
\end{enumerate}
}
\proof{See \cite[Proposition 2.4]{BlanesStability}. \qedhere}\medskip

Before determining the stability threshold $z_*$ for a given force-gradient integrator \eqref{eq:force-gradient-integrator}, we define an upper bound for $z_*$.

\definition{Let $\mathfrak{p}(z)$ be the stability polynomial of a consistent force-gradient integrator \eqref{eq:force-gradient-integrator}. We define $z^*$ as the largest real non-negative number such that $\lvert \mathfrak{p}(z)\rvert \le 1\; \forall z \in [0,z^*]$.}\normalfont\medskip

\remark{
It is clear from the proof of Proposition 1 that $[-z^*,z^*]$ is the largest interval including $0$ such that $\stabilitymatrix(z)$ has eigenvalues of modulus one and therefore $z_* \le z^*$. 
}\normalfont\medskip
For a consistent self-adjoint force-gradient integrator \eqref{eq:force-gradient-integrator}, it holds $\mathfrak{p}(z) = 1-\frac{z^2}{2} + \mathcal{O}(z^4)$ and thus $\mathfrak{p}''(0) < 0$. Then, $\mathfrak{p}(z)^2 -1 \le 0$ for sufficiently small $\lvert z \rvert$, and in that case $z^*$ is the smallest real positive zero with odd multiplicity of the polynomial $\mathfrak{p}(z)^2 - 1$.
Therefore, potential instabilities may occur within the interval $[0,z^*]$ at the real zeros with even multiplicity of the polynomial $\mathfrak{p}(z)^2 -1$. 
The following proposition will determine the stability threshold $z_*$.

\proposition{Let $\stabilitymatrix(z)$ denote the stability matrix of a self-adjoint force-gradient integrator \eqref{eq:force-gradient-integrator} and $\mathfrak{p}(z)$ its stability polynomial. 
Suppose that $0 = z_0 < z_1 < \ldots < z_{m}$ are the real zeros with even multiplicity of the polynomial $\mathfrak{p}(z)^2 - 1$ in the interval $[0,z^*]$. Then $z_* = z^*$ if
\begin{equation}\label{eq:stability_condition}
    K_{2,1}(z_\ell) = K_{1,2}(z_\ell) = 0
\end{equation}
for each $\ell=1,\ldots,m$. Otherwise, $z_*$ is the smallest $z_\ell$ that violates condition \eqref{eq:stability_condition}.
}
\proof{See \cite[Proposition 2.7]{BlanesStability}. \qedhere}\normalfont\medskip

\example{We again consider the force-gradient integrator \eqref{eq:BACAB}. 
The polynomial $\mathfrak{p}(z)^2 - 1$ has no zero with even multiplicity except $z_0 = 0$. Thus, the force-gradient integrator achieves the optimal stability threshold $z_* = z^* = 2 \sqrt{3}$.
}\normalfont%

\section{Detection of efficient decomposition algorithms}
\label{sec:detection}
The integrator coefficients of decomposition algorithms can be selected to achieve the maximum attainable convergence order.
For many versions, the order conditions establish an under-determined system, leaving certain integrator coefficients as degrees of freedom.
Although selecting the degrees of freedom does not alter the convergence order $\convergenceorder$, it does influence the accuracy and the stability threshold $z_*$ of the integrator.
A commonly employed procedure \cite{mclachlan1995,omelyan2003symplectic} is to identify the integrator coefficients that minimize a given norm of the leading error coefficients.

\subsection{Minimum-error methods}
Given a decomposition algorithm of convergence order $\convergenceorder$, the integrator possesses a local truncation error of order $\mathcal{O}(h^{\convergenceorder+1})$, which is a linear combination of nested commutators whose coefficients are determined by the violations in the order conditions for convergence order $\convergenceorder + 1$.
For instance, applying a self-adjoint decomposition algorithm $\Phi_{\stepsize}$ of convergence order $\convergenceorder = 2$ to the Hamiltonian system \eqref{eq:Hamiltonian_system} yields a local truncation error of the form
\begin{align}\label{eq:local_truncation_error_2nd_order} 
\begin{split}
\Phi_{\stepsize}(\q_0,\p_0) &- (\q(\stepsize),\p(\stepsize)) \\
=\ &\stepsize^3 \left( \alpha [\Kineticvf,[\Kineticvf,\Potentialvf]] + \beta [\Potentialvf,[\Kineticvf,\Potentialvf]] \right) + \mathcal{O}(\stepsize^5), 
\end{split}
\end{align}
with leading error coefficients $\alpha,\beta \in \mathbb{R}$, $\alpha \cdot \beta \neq 0$.
Minimizing the principal error term will result in a more accurate numerical integration scheme. Since the magnitudes of the commutators are problem-specific, a common approach is to equate all commutators to one and then minimize some norm of the leading error coefficients. Typically, the Euclidean norm is chosen. 
For our example \eqref{eq:local_truncation_error_2nd_order}, this implies that we seek to minimize the expression
\begin{equation}\label{eq:norm-to-minimize_order2}
    \sqrt{\alpha^2 + \beta^2}.
\end{equation}
For higher orders, this concept can be extended in a straightforward manner; see for example~\cite{omelyan2003symplectic,schäfers2024hessianfree}.
Additionally, the computational cost must be taken into account to determine which integrator is more efficient, i.e., which integrator achieves the desired accuracy at the least computational cost.
In lattice QCD, the computational cost is governed by the evaluations of the forces and the force-gradient terms. 
Consequently, a common way to compare the efficiency of two decomposition algorithms of the same order $\convergenceorder$ is given by the efficiency measure~\cite{omelyan2003symplectic,schäfers2024hessianfree} 
\begin{equation}\label{eq:EffErr}
    \EffErr{\convergenceorder} = \frac{1}{(n_f + \xi n_g)^\convergenceorder \cdot \mathrm{Err}_{n+1}},
\end{equation}
where $n_f$ and $n_g$ denote the number of force evaluations and force-gradient evaluations per time step, respectively. Furthermore, $\xi > 0$ is a constant that quantifies the relative cost difference between force-gradient and force evaluations. $\mathrm{Err}_{n+1}$ denotes a (possibly weighted) norm of the leading error coefficients, as exemplified in~\eqref{eq:norm-to-minimize_order2}.
This approach has been employed to derive self-adjoint minimum-error methods with $s \leq 6$ stages.
On the one hand, in~\cite{omelyan2003symplectic}, the derivation of minimum-error force-gradient integrators was conducted by selecting $\xi = 2$ and employing the non-weighted Euclidean norm for the principal error term.
On the other hand, in~\cite{schäfers2024hessianfree}, we conducted similar investigations for Hessian-free force-gradient integrators by setting $\xi = 1$ (as the force-gradient evaluation is replaced by another force evaluation) and a modified Euclidean norm that incorporates additional error terms with appropriate weights.
Both investigations include conventional splitting methods as a special case.

\subsection{The relative stability threshold}
\begin{figure}[htb]
    \centering
    \begin{subfigure}[b]{\columnwidth}
        \centering
        \includegraphics{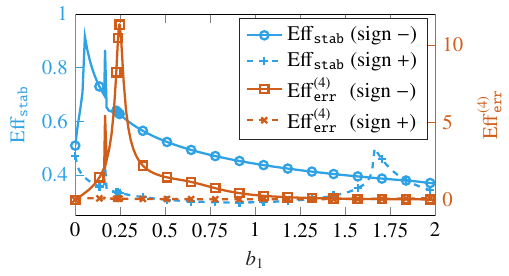} %
        \caption{Results for the force-gradient integrator ABACABA \eqref{eq:ABACABA}. For $b_1 = 1/6$, the method reduces to BACAB \eqref{eq:BACAB}, explaining the respective peak.}
        \label{fig:ABACABA_analysis}
    \end{subfigure}

    \begin{subfigure}[b]{\columnwidth}
        \centering
        \includegraphics{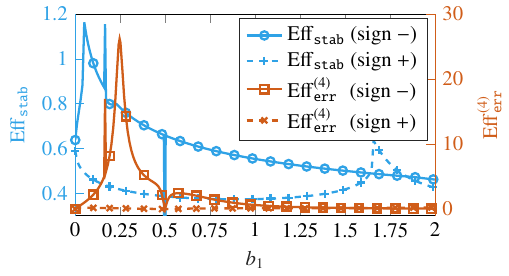} %
        \caption{Results for the Hessian-free force-gradient integrator ABADABA \eqref{eq:ABADABA}. For $b_1 = 1/6$, the method reduces to BADAB \eqref{eq:BADAB}, whereas for $b_1 = 1/2$, we have $b_2 = 0$ so that the approximated force-gradient update \eqref{eq:modified_momentum_update} is not defined, explaining the respective peaks.}
        \label{fig:ABADABA_analysis}
    \end{subfigure}
    \caption{The efficiency measures $\EffErr{4}$ and $\EffStab$ for the decomposition algorithms \eqref{eq:ABAXABA} evaluated at different values of $b_1 \in (0,2]$.}
    \label{fig:ABAXABA_analysis}
\end{figure}

The investigation of the linear stability of (Hessian-free) force-gradient integrators presents an alternative approach to identifying promising decomposition algorithms.
By investigating the stability threshold, we can determine which integrator enables stable integration at the lowest computational cost.
Consider two integrators, $\Phi_{\stepsize}^{\{1\}}$ and $\Phi_{\stepsize}^{\{2\}}$, with corresponding stability thresholds, $z_*^{\{1\}}$ and $z_*^{\{2\}}$, and computational costs per time step, $n_f^{\{1\}} + \xi n_g^{\{1\}}$, and $n_f^{\{2\}} + \xi n_g^{\{2\}}$, respectively. 
For a given $\omega > 0$, the maximum permissible step size ensuring the stability of the first integrator is $\stepsize_{\mathrm{max}}^{\{1\}} = z_*^{\{1\}}/\omega$. 
To achieve the same computational cost as $\Phi^{\{1\}}_{\stepsize_{\mathrm{max}}^{\{1\}}}$, the second integrator must be employed with the step size 
$$\tilde{\stepsize} = \frac{z_*^{\{1\}}}{\omega} \cdot \frac{n_f^{\{2\}} + \xi n_g^{\{2\}}}{n_f^{\{1\}} + \xi n_g^{\{1\}}}.$$
If $\stepsize_{\mathrm{max}}^{\{2\}} = z_*^{\{2\}}/\omega < \tilde{\stepsize}$,
then $\Phi_{\tilde{\stepsize}}^{\{2\}}$ is unstable, implying that $\Phi^{\{1\}}_{\stepsize_{\mathrm{max}}^{\{1\}}}$ enables a stable integration at a lower computational cost. 
Consequently, $\Phi^{\{1\}}_{\stepsize}$ is computationally more efficient than $\Phi^{\{2\}}_{\stepsize}$, provided that
$$\frac{z_*^{\{2\}}}{\omega} < \frac{z_*^{\{1\}}}{\omega} \cdot \frac{n_f^{\{2\}} + \xi n_g^{\{2\}}}{n_f^{\{1\}} + \xi n_g^{\{1\}}} \quad \Leftrightarrow \quad \frac{z_*^{\{2\}}}{n_f^{\{2\}} + \xi n_g^{\{2\}}} < \frac{z_*^{\{1\}}}{n_f^{\{1\}} + \xi n_g^{\{1\}}},$$
i.e., we seek to maximize the efficiency measure
\begin{equation}\label{eq:EffStab}
    \EffStab = z_* / (n_f + \xi n_g),
\end{equation}
which we refer to as the \emph{relative stability threshold}.
We conducted a comprehensive analysis of all variants of self-adjoint (Hessian-free) force-gradient integrators with $s \leq 6$ stages.
Exemplarily, we present results for a variant with one degree of freedom and a variant with two degrees of freedom. 
Firstly, Figure~\ref{fig:ABAXABA_analysis} depicts the results for the force-gradient integrator ABACABA
\begin{subequations}\label{eq:ABAXABA}
\begin{align}\label{eq:ABACABA}
\begin{split}
    \Phi_\stepsize = &\e^{a_1 \stepsize \Kineticvf} \e^{b_1 \stepsize \Potentialvf} \e^{(0.5 - a_1)\stepsize \Kineticvf} \e^{(1-2b_1)\stepsize \Potentialvf + c_2 \stepsize^3 \ForceGradientvf} \\
    &\quad\e^{(0.5 - a_1)\stepsize \Kineticvf} \e^{b_1 \stepsize \Potentialvf} \e^{a_1 \stepsize \Kineticvf}
\end{split}
\intertext{and its Hessian-free variant ABADABA} 
\begin{split}
    \label{eq:ABADABA}
    \Phi_\stepsize = &\e^{a_1 \stepsize \hat{\mathcal{T}}} \e^{b_1 \stepsize \hat{\mathcal{V}}} \e^{(0.5 - a_1)\stepsize \hat{\mathcal{T}}} \e^{(1-2b_1)\stepsize \hat{\mathcal{D}}(\stepsize,1-2b_1,c_2)} \\
    &\quad\e^{(0.5 - a_1)\stepsize \hat{\mathcal{T}}} \e^{b_1 \stepsize \hat{\mathcal{V}}} \e^{a_1 \stepsize \hat{\mathcal{T}}}
\end{split}
\end{align}
with integrator coefficients
\begin{equation}\label{eq:ABAXABA_coeffs}
    a_1 = \frac{1}{2} \pm \frac{1}{\sqrt{24b_1}}, \qquad c_2 = \frac{1}{12} \left( 1 \pm \sqrt{6b_1}(1-b_1) \right), 
\end{equation}
\end{subequations}
and one degree of freedom $b_1 > 0$.
On the one hand, the results demonstrate that the measures exhibit comparable behavior for both conventional force-gradient integrators and their Hessian-free counterparts.
On the other hand, it is evident that the values for $\EffErr{4}$ are already experiencing a substantial decline for minor deviations from the minimum-error variant. Conversely, the relative stability threshold $\EffStab$ of the minimum-error method remains reasonably high in comparison to the maximum attainable value.

\begin{figure}[htb!]
    \centering
    \begin{subfigure}[t]{\columnwidth} 
        \centering
        \includegraphics[scale=.97125]{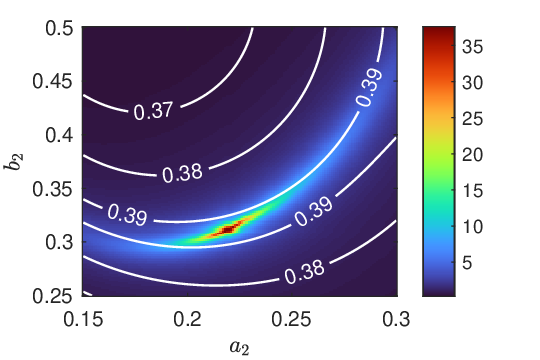}
        \caption{Force-gradient integrator BACABACAB \eqref{eq:BACABACAB}.}
        \label{fig:BACABACAB_analysis}
    \end{subfigure}
    \hfill 
    \begin{subfigure}[t]{\columnwidth}
        \centering
        \includegraphics[scale=.97125]{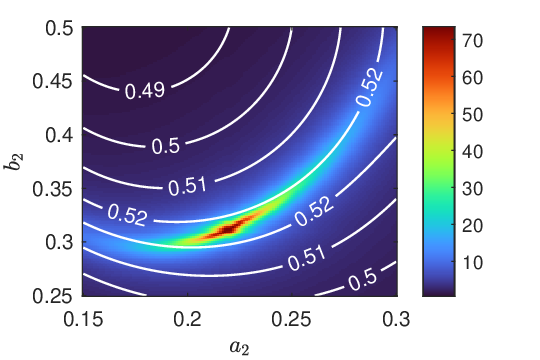}
        \caption{Hessian-free force-gradient integrator BADABADAB \eqref{eq:BADABADAB}.}
        \label{fig:BADABADAB_analysis}
    \end{subfigure}
    \caption{The efficiency measure $\EffErr{4}$ for the decomposition algorithms \eqref{eq:BAXABAXAB}
    evaluated at different values of $a_2 \in [0.15,0.3]$ and $b_2 \in [0.25,0.5]$. Moreover, the white contour lines indicate the values for the relative stability threshold $\EffStab$.}
    \label{fig:BAXABAXAB_analysis}
\end{figure}

We obtain comparable results for multiple degrees of freedom. For instance, the results for the force-gradient integrator BACABACAB 
\begin{subequations}\label{eq:BAXABAXAB}
\begin{align}\label{eq:BACABACAB}
\begin{split}
     \Phi_\stepsize = &\e^{b_1 \stepsize \Potentialvf} \e^{a_2 \stepsize \Kineticvf} \e^{b_2 \stepsize \Potentialvf + c_2 \stepsize^3 \ForceGradientvf} \e^{(0.5-a_2) \stepsize \Kineticvf} \e^{(1 - 2(b_1+b_2)) \stepsize \Potentialvf} \\
     &\quad\e^{(0.5-a_2) \stepsize \Kineticvf} \e^{b_2 \stepsize \Potentialvf + c_2 \stepsize^3 \ForceGradientvf} \e^{a_2 \stepsize \Kineticvf} \e^{b_1 \stepsize \Potentialvf}
\end{split}
\intertext{and its Hessian-free counterpart BADABADAB} 
    \label{eq:BADABADAB}
\begin{split}
     \Phi_\stepsize = &\e^{b_1 \stepsize \Potentialvf} \e^{a_2 \stepsize \Kineticvf} \e^{b_2 \stepsize \hat{\mathcal{D}}(\stepsize,b_2,c_2)} \e^{(0.5-a_2) \stepsize \Kineticvf} \e^{(1 - 2(b_1+b_2)) \stepsize \Potentialvf} \\
     &\quad\e^{(0.5-a_2) \stepsize \Kineticvf} \e^{b_2 \stepsize \hat{\mathcal{D}}(\stepsize,b_2,c_2)} \e^{a_2 \stepsize \Kineticvf} \e^{b_1 \stepsize \Potentialvf}
\end{split}
\end{align}
with integrator coefficients
\begin{align}\label{eq:BAXABAXAB_coeffs}
\begin{split}
    b_1 &= - 4 b_2 a_2^2 + 4 b_2 a_2 - b_2 + \frac{1}{6}, \\
    c_2 &= 4 a_2^4 b_2^2 - 4 a_2^3 b_2^2 + \frac{2 a_2^2 b_2}{3} + \frac{a_2 b_2^2}{2} - \frac{a_2 b_2}{3} + \frac{1}{144},
\end{split}
\end{align}
\end{subequations}
and two degrees of freedom $a_2,b_2 \in \mathbb{R}$ are depicted in Figure~\ref{fig:BAXABAXAB_analysis}. 
Once again, both the conventional force-gradient integrator and the Hessian-free variant exhibit similar behavior. Furthermore, the results underscore the sensitivity of the efficiency measure $\EffErr{4}$ to deviations from the minimum-error coefficients derived in \cite{omelyan2003symplectic} and \cite{schäfers2024hessianfree}, respectively.
Theoretically, it is possible to enhance the relative stability threshold $\EffStab$ by approximately a factor of two by selecting integrator coefficients outside the depicted region.
In contrast, within these stability-enhanced regions, the efficiency measure $\EffErr{4}$ achieves remarkably low values. Consequently, a significantly smaller step size is required to attain the same level of accuracy as observed in the minimum-error variant.
In comparison to \eqref{eq:BAXAB}, these stability-enhanced versions exhibit smaller values for both $\EffErr{4}$ and $\EffStab$, rendering them impractical for applications.

For other variants of (Hessian-free) force-gradient integrators with $s \leq 6$ stages, we observe a similar behavior. 

\subsection{Detection of efficient decomposition algorithms}
For a complete classification, we identified non-dominated integrator variants of convergence order $\convergenceorder \in \{2,4,6\}$, which guarantee that no other variant of the same convergence order achieves simultaneously higher values for both efficiency measures \EffErr{\convergenceorder} and \EffStab.

\subsubsection{Convergence order two}
\begin{figure}[tbh]
    \centering
    \includegraphics{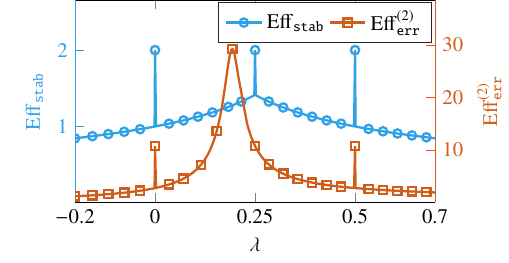}
    \caption{Visualization of $\EffErr{2}(\lambda)$ \eqref{eq:Eff_err_OMF2} and $\EffStab(\lambda)$ \eqref{eq:Eff_stab_OMF2} for the splitting methods BABAB \eqref{eq:BABAB} and ABABA \eqref{eq:ABABA}.}
    \label{fig:OMF2_analysis}
\end{figure}

For convergence order $\convergenceorder = 2$, conventional splitting methods prove to be the most efficient choice. Notably, the Strang \cite{strang1968construction} splittings BAB
\begin{subequations}\label{eq:Strang-splitting}
\begin{align}
    \Phi_\stepsize &= \e^{\tfrac{\stepsize}{2} \Potentialvf} \e^{\stepsize \Kineticvf} \e^{\tfrac{\stepsize}{2} \Potentialvf} \label{eq:BAB}
\intertext{and ABA}
\label{eq:ABA}
    \Phi_\stepsize &= \e^{\tfrac{\stepsize}{2} \Kineticvf} \e^{\stepsize \Potentialvf} \e^{\tfrac{\stepsize}{2} \Kineticvf},
\end{align}
\end{subequations}
as well as the $(s=3)$-stage splitting methods BABAB
\begin{subequations}\label{eq:OMF2}
\begin{align}
\label{eq:BABAB}
    \Phi_\stepsize &= \e^{\lambda \stepsize \Potentialvf} \e^{\tfrac{\stepsize}{2} \Kineticvf} \e^{(1-2\lambda)\stepsize \Potentialvf} \e^{\tfrac{\stepsize}{2} \Kineticvf} \e^{\lambda\stepsize \Potentialvf}
\intertext{and ABABA}
\label{eq:ABABA}
    \Phi_\stepsize &= \e^{\lambda \stepsize \Kineticvf} \e^{\tfrac{\stepsize}{2} \Potentialvf} \e^{(1-2\lambda)\stepsize \Kineticvf} \e^{\tfrac{\stepsize}{2} \Potentialvf} \e^{\lambda \stepsize \Kineticvf}
\end{align}
\end{subequations}
with one degree of freedom $\lambda \in \mathbb{R}$, are promising variants. 
For the Strang splittings \eqref{eq:Strang-splitting}, we obtain efficiency values of $\EffErr{2} \approx 10.73$ and $\EffStab = 2$.
Note that a relative stability threshold of $\EffStab = 2$ is the maximum possible value for decomposition algorithms. 
For the integrators \eqref{eq:OMF2}, we obtain
\begin{align}\label{eq:Eff_err_OMF2}
    \EffErr{2}(\lambda) &= \frac{6}{(144\lambda^4 - 288 \lambda^3 + 228 \lambda^2 - 60 \lambda + 5)^{1/2}}, \\
    \EffStab(\lambda) &= 
    \begin{cases}
        1/(1-2\lambda)^{1/2}, & \lambda \in \{ x \in (-\infty,\tfrac{1}{4}) \,\vert\, x \neq 0 \}, \\
        1/(2\lambda)^{1/2}, & \lambda \in \{x \in (\tfrac{1}{4},\infty) \,\vert\, x \neq \tfrac{1}{2}\}, \\
        1/(1-2\lambda)^{1/2}, & \lambda = 0, \\
        2, & \lambda = \tfrac{1}{4}, \\
        2/(2\lambda)^{1/2}, & \lambda = \tfrac{1}{2}.
    \end{cases} \label{eq:Eff_stab_OMF2}
\end{align}
A visualization of both efficiency measures is presented in Figure~\ref{fig:OMF2_analysis}. 
The minimum-error variant with $\EffErr{2} \approx 29.24$ is obtained for~\cite{mclachlan1995,omelyan2003symplectic}
\begin{align}\label{eq:OMF2_min-error_coeff}
\begin{split}
    \lambda_{\mathtt{err}} &= \frac{1}{2} - \frac{(2 \sqrt{326} + 36)^{1/3}}{12} + \frac{1}{6(2 \sqrt{326} + 36)^{1/3}} \\
    &\approx 0.1931833275037836.
\end{split}
\end{align}
Conversely, the relative stability threshold is maximized for $\lambda_{\mathtt{stab}} = 0.25$, where the variants reduce to two applications of the respective Strang splitting. 
We also obtain \mbox{$\EffStab = 2$} for $\lambda \in \{0, 0.5\}$ because, for these values, the splitting methods \eqref{eq:OMF2} reduce to BAB and ABA, respectively.
The interval $\lambda \in [\lambda_{\mathtt{err}},\lambda_{\mathtt{stab}}]$ encompasses all non-dominant variants of the integrators \eqref{eq:OMF2}.
Depending on the specific ensemble, selecting a value within this interval may lead to the most efficient computational process by providing the optimal trade-off between accuracy and numerical stability.
Particularly, for lighter fermion masses, a value $\lambda > \lambda_{\mathtt{err}}$ should be preferred.
However, it should be noted that, depending on the specific ensemble, the Poisson brackets may be vastly different in magnitude so that the efficiency measure $\EffErr{\convergenceorder}$ is not very accurate. Using appropriate weights for the Poisson brackets and minimizing a weighted norm instead will then result in a different optimal value $\lambda_{\mathtt{err}}$.

\subsubsection{Convergence order four}
Most variants of (Hessian-free) force-gradient integrators with $s\leq 6$ stages exhibit convergence order $\convergenceorder = 4$.
In \cite{omelyan2003symplectic}, there are 24 different variants of conventional force-gradient integrators of order $\convergenceorder = 4$, whereas the Hessian-free framework has 28 different variants with $s \leq 6$ stages and $\convergenceorder = 4$ \cite{schäfers2024hessianfree}.
Moreover, there are 6 different variants of conventional splitting methods with $\convergenceorder = 4$ and $s \leq 6$.
As these integrators have up to four degrees of freedom in the integrator coefficients, it becomes a computationally demanding task to explore the entire family of solutions.
To obtain numerical evidence, we evaluated both measures $\EffErr{4}$ and $\EffStab$ on a discrete grid with values $a_j,b_j \in [-2,2]$ and $c_j \in [-0.1,0.1]$ for all degrees of freedom.

Considering the framework of conventional force-gradient integrators, we find five non-dominated integrator variants of order $\convergenceorder = 4$ with $s \leq 6$ stages. These integrators are listed in Table~\ref{tab:order4_overview_FGI}. 
It is noteworthy that three of the five integrators are conventional splitting methods, while the force-gradient integrators are particularly interesting due to their high accuracy at the cost of less numerical stability. 
However, it is important to note that the force-gradient integrator BACAB \eqref{eq:BACAB} exhibits very similar values to BABABABAB \eqref{eq:BABABABAB} (see Table~\ref{tab:force-gradient_overview}). 
For less expensive evaluations of the force-gradient term \eqref{eq:force-gradient_term}, $\xi < 2$, the force-gradient integrator \eqref{eq:BACAB} exhibits larger values for both efficiency measures, making it a promising choice that is particularly advantageous in terms of the relative stability threshold $\EffStab$.
Furthermore, the variants ABACABA \eqref{eq:ABACABA} and BACABACAB \eqref{eq:BACABACAB} are dominated by the splitting method BABABABABAB due to the choice $\xi = 2$. For $\xi < 2$, the aforementioned variants may turn out as non-dominated variants as well.

In contrast, for the Hessian-free framework, we identified six non-dominated integrators, which are summarized in Table~\ref{tab:order4_overview_HfFGI}. 
Notably, five of these integrators are indeed Hessian-free force-gradient integrators, highlighting their efficiency in terms of both accuracy and numerical stability. 
The integrator BADABABADAB \eqref{eq:BADABABADAB} is a non-dominated variant, but it exhibits a significantly smaller value for $\EffErr{4}$ and only a slightly greater value for $\EffStab$ compared to ABADABADABA \eqref{eq:ABADABADABA}. Therefore, preference should be given to ABADABADABA.
It is important to note that we did not compare conventional force-gradient integrators with Hessian-free force-gradient integrators. The efficiency of these integrators depends on the specific value of $\xi$ in \eqref{eq:EffErr} and whether the force-gradient term is available in the software implementation.

For both frameworks, it is important to highlight that it is possible to attain slightly improved values for $\EffStab$ by deviating slightly from the minimum-error coefficients. However, these alterations entail a substantial reduction in values for $\EffErr{4}$, as illustrated, for instance, in Figure~\ref{fig:BAXABAXAB_analysis}.
Due to the substantial reduction in accuracy, these alternative non-dominant variants are unlikely to lead to a more efficient computational process.
Consequently, we propose to continue optimizing the efficiency measure $\EffErr{4}$ and select one of the non-dominated variants listed in Table~\ref{tab:order4_overview_FGI} and Table~\ref{tab:order4_overview_HfFGI}, respectively.
Selecting a variant with a higher value for $\EffErr{4}$ corresponds to a smaller value for $\EffStab$, and vice versa.
Depending on the specific problem, one variant will provide the optimal trade-off between accuracy and stability, resulting in the most efficient computational process.

\begin{table*}[tbh!]
    \centering
    \begin{tabular}{@{}l c c l r c r c@{}}
        \toprule 
        Integrator & $n_f$ & $n_g$ & $\mathrm{Err}_{5}$ & $\EffErr{4}$ & $z_*$ & $\EffStab$ & Equations \\
        \midrule 
        BABABABAB & 4 & 0 & 0.000654 & 5.97 & 3.4696 & 0.8674 & \eqref{eq:BABABABAB} \\
        ABABABABA & 4 & 0 & 0.000610 & 6.40 & 2.9894 & 0.7474 & \eqref{eq:ABABABABA} \\
        BABABABABAB & 5 & 0 & 0.0000270 & 59.26 & 3.1421 & 0.6284 & \eqref{eq:BABABABABAB} \\
        CABACABAC & 4 & 2 & 0.00000368 & 66.34 & 3.0883 & 0.3860 & \eqref{eq:CABACABAC} \\
        ABACABACABA & 5 & 2 & 0.00000127 & 120.03 & 3.1223 & 0.3469 & \eqref{eq:ABACABACABA} \\
        \bottomrule
    \end{tabular}
    \caption{Overview of non-dominated fourth-order conventional force-gradient integrators with $s \leq 6$ stages. The values refer to the minimum-error variants that have been derived in \cite{omelyan2003symplectic}.}
    \label{tab:order4_overview_FGI}
\end{table*}

\begin{table*}[tbh!]
    \centering
    \begin{tabular}{@{}l c c l r c r c@{}}
        \toprule 
        Integrator & $n_f$ & $n_g$ & $\mathrm{Err}_{5}$ & $\EffErr{4}$ & $z_*$ & $\EffStab$ & Equations \\
        \midrule 
        BADAB & 2 & 1 & 0.000728 & 16.96 & 3.4641 & 1.1547 & \eqref{eq:BADAB} \\
        ABADABA & 3 & 1 & 0.0000149 & 26.19 & 3.1377 & 0.7844 & \eqref{eq:ABADABA}, \eqref{eq:ABADABA_min-error} \\
        BABABABABAB & 5 & 0 & 0.0000270 & 59.26 & 3.1421 & 0.6284 & \eqref{eq:BABABABABAB} \\
        BADABADAB  & 4 & 2 & 0.0000105 & 73.45 & 3.1457 & 0.5243 & \eqref{eq:BADABADAB}, \eqref{eq:BADABADAB_min-error}  \\
        BADABABADAB & 5 & 2 & 0.00000520 & 80.13 & 3.1371 & 0.4482 & \eqref{eq:BADABABADAB} \\
        ABADABADABA  & 5 & 2 & 0.00000445 & 93.60 & 3.1239 & 0.4463 & \eqref{eq:ABADABADABA} \\        
        \bottomrule
    \end{tabular}
    \caption{Overview of non-dominated fourth-order Hessian-free force-gradient integrators with $s \leq 6$ stages. The values refer to the minimum-error variants that have been derived in \cite{schäfers2024hessianfree}.}
    \label{tab:order4_overview_HfFGI}
\end{table*}

\subsubsection{Convergence order six}
The minimum-error variant of the 8-stage velocity version BABABABABABABAB \cite{omelyan2003symplectic}
\begin{subequations}\label{eq:BABABABABABABAB}
\begin{align}
\begin{split}
    \Phi_\stepsize = &\e^{b_1 \stepsize \Potentialvf} \e^{a_2 \stepsize \Kineticvf} \e^{b_2 \stepsize \Potentialvf} \e^{a_3 \stepsize \Kineticvf} \e^{b_3 \stepsize \Potentialvf} \e^{a_4 \stepsize \Kineticvf}\\
    &\quad\e^{\left(\tfrac{1}{2} -(b_1+b_2+b_3)\right) \stepsize \Potentialvf} \e^{(1 - 2(a_2 + a_3 + a_4)) \stepsize \Kineticvf} \\
    &\qquad \e^{\left(\tfrac{1}{2} -(b_1+b_2+b_3)\right) \stepsize \Potentialvf} \e^{a_4 \stepsize \Kineticvf} \e^{b_3 \stepsize \Potentialvf} \e^{a_3 \stepsize \Kineticvf} \\
    &\qquad\quad \e^{b_2 \stepsize \Potentialvf} \e^{a_2 \stepsize \Kineticvf} \e^{b_1 \stepsize \Potentialvf} 
\end{split}
\end{align}
with coefficients
\begin{equation}
\begin{aligned}
    a_2 &= \phantom{-}0.2465881872786138, \\
    a_3 &= \phantom{-}0.6047073875057809, \\
    a_4 &= -0.4009869039788007, \\
    b_1 &= \phantom{-}0.0833333333333333, \\
    b_2 &= \phantom{-}0.3977675859548440, \\
    b_3 &= -0.0393336931446257, 
\end{aligned}
\end{equation}
\end{subequations}
and $\EffErr{6} \approx 1.40$, $\EffStab \approx 0.4515$, turns out to dominate all sixth-order self-adjoint (Hessian-free) force-gradient integrators with $s \leq 6$ stages (see Table~\ref{tab:force-gradient_overview} and Table~\ref{tab:Hessian-free_overview} for a comparison to the other variants). 
The investigations in \cite{omelyan2003symplectic,schäfers2024hessianfree} only consider (Hessian-free) force-gradient integrators with \mbox{$s \leq 6$} stages. For \mbox{$s > 6$}, there exist more variants of sixth-order integrators that may offer additional non-dominated alternatives. However, these variants have not yet been investigated.

\subsubsection{Remarks}
Although investigations in \cite{omelyan2003symplectic} and \cite{schäfers2024hessianfree} did not explicitly address numerical stability, it is evident that maximizing the efficiency measure $\EffErr{\convergenceorder}$, resulting in minimum-error variants, yields promising integrator variants that provide the optimal trade-off between accuracy and numerical stability.
By incorporating the proposed relative stability threshold $\EffStab$ into these investigations, it is possible to identify efficient integrator variants that excel in both accuracy and numerical stability.
 
For convergence order $\convergenceorder = 4$, in particular, the findings demonstrate remarkable efficiency of the force-gradient approach.
For convergence order $\convergenceorder = 6$, the family of (Hessian-free) force-gradient integrators with $s \leq 6$ stages fails to yield any variant that surpasses the splitting method \eqref{eq:BABABABABABABAB}. The family of (Hessian-free) force-gradient integrators with $s \geq 7$ stages may potentially yield promising variants, which require further investigation.
 
For completeness, the definitions of all non-dominated variants that have not yet been introduced are given in~\ref{app:Integrators}. Moreover, \ref{app:Integrators} provides a summary of the results for all minimum-error variants of (Hessian-free) force-gradient integrators with $s \leq 6$ stages, leading to extended versions of the tables \cite[Table 2]{omelyan2003symplectic} and \cite[Table 1]{schäfers2024hessianfree}.

\section{Numerical tests}
\label{sec:numerical_tests}
We will conduct numerical tests demonstrating that incorporating the relative stability threshold $\EffStab$ in our investigations results in an enhanced explainability of the computational efficiency of decomposition algorithms when applied to lattice QCD simulations.
Particularly, in Section \ref{sec:Schwinger}, we investigate potential differences in the numerical stability of a conventional force-gradient integrator and its Hessian-free variant when applied to the two-dimensional Schwinger model.
Subsequently, in Section \ref{sec:Em1}, we will extend numerical tests from \cite{schäfers2024hessianfree,Schäfers:2025BT} with two heavy Wilson fermions and demonstrate that the relative stability threshold $\EffStab$ enables the explanation of the efficiency of the numerical integration schemes.
Finally, in Section \ref{sec:tm-simulations}, we perform twisted mass simulations and provide numerical evidence that the stability threshold $z_*$ provides a reliable stability criterion for lattice QCD simulations.

\subsection{The two-dimensional Schwinger model}\label{sec:Schwinger}
\begin{figure}[htb!]
    \centering
    \includegraphics{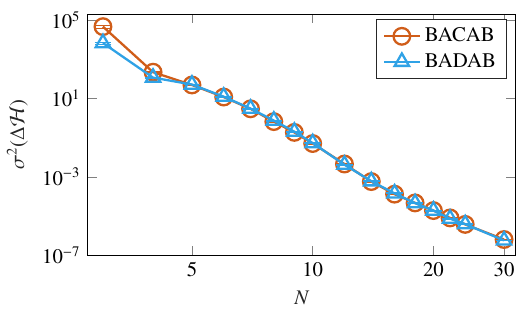}
    \caption{Two-dimensional Schwinger model. Variance of $\Delta\Hamiltonian$ vs.~the number of time steps per trajectory $N$ for the conventional force-gradient integrator BACAB \eqref{eq:BACAB} ($\circ$) and its Hessian-free counterpart BADAB \eqref{eq:BADAB} ($\triangle$). For any $N$, $1000$ trajectories of length $\tau = 4.0$ have been computed.}
    \label{fig:Schwinger_results}
\end{figure}

The linear stability analysis for force-gradient integrators and their Hessian-free variants coincides, as both frameworks are equivalent when applied to linear ODE systems, such as the harmonic oscillator~\eqref{eq:harmonic_oscillator}. 
However, when applied to lattice QCD simulations, the two frameworks differ. 
Consequently, the natural question arises regarding the impact of the approximation \eqref{eq:modified_momentum_update} on the numerical stability of the integrator. 
The two-dimensional Schwinger model shares many of the features of QCD simulations. Moreover, for this system, the force-gradient term has been implemented \cite{shcherbakov2017adapted}, making it an ideal test problem to investigate whether approximating the force-gradient update via \eqref{eq:modified_momentum_update} affects the numerical stability of the integrator.
As an example, we compare the numerical stability of the conventional force-gradient integrator BACAB \eqref{eq:BACAB} and its Hessian-free variant BADAB \eqref{eq:BADAB} by computing $1000$ trajectories of length $\tau = 4.0$ on a $16\times 16$ lattice with parameters $\beta=1.0$ and $m=0.352443$ (see \cite{Christian:2005yp}) for different number of time steps $N$ per trajectory. 
The results in Figure~\ref{fig:Schwinger_results} demonstrate that both integrators become unstable simultaneously. Consequently, we have numerical evidence suggesting that both frameworks possess similar numerical stability properties when employed to lattice field theories. This aligns with the linear stability analysis, where both frameworks are equivalent.
Moreover, both frameworks yield comparable values for $\sigma^2(\Delta\Hamiltonian)$ when employing the same number of time steps $N$. 
Given the lower computational cost of the Hessian-free variant, it is the more efficient choice.

\subsection{Lattice QCD simulations with two heavy Wilson fermions}\label{sec:Em1} 
We consider an ensemble used in~\cite{Knechtli2022} generated with two dynamical non-perturbatively $\mathcal{O}(a)$ improved Wilson quarks at a mass equal to half of the physical charm quark. 
The fermion part is decomposed using even-odd reduction in combination with one Hasenbusch mass preconditioning term \cite{Hasenbusch2001} with shift parameter $\mu$. For a more detailed discussion of the decomposition, we refer to \cite{Frommer2014}. 
The numerical tests are performed on a $48 \times 24^3$ lattice with gauge coupling $\beta = 5.3$ and hopping parameter \mbox{$\kappa = 0.1327$} using an extended version of \texttt{openQCD v2.4}\footnote{\href{https://github.com/KevinSchaefers/openQCD_force-gradient}{https://github.com/KevinSchaefers/openQCD\_force-gradient}.} including the framework of Hessian-free force-gradient integrators by putting all forces on a single time scale of integration.
This setup has already been investigated in the initial work on Hessian-free force-gradient integrators \cite{schäfers2024hessianfree} where it was observed that the efficiency measure $\mathrm{Eff}_{\mathtt{err}}^{(\convergenceorder)}$ does not perfectly match with the performance of the numerical integration schemes.

\begin{figure*}[ht!]
    \centering
    \begin{subfigure}[b]{0.49\textwidth}
        \centering
        \includegraphics{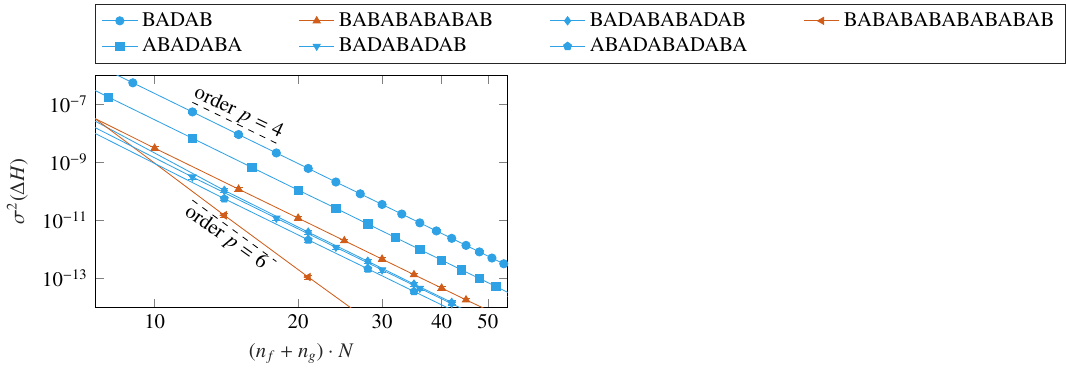}
        \caption{Trajectory length $\tau=0.1$. The second-order integrators BAB \eqref{eq:BAB} and BABAB \eqref{eq:BABAB} achieve values of $\sigma^2(\Delta \Hamiltonian) > 10^{-7}$ for $(n_f + n_g)\cdot N = 50$ and are thus not included for visualization purposes.}
        \label{fig:Em1_scaling-phase}
    \end{subfigure}
    \hfill
    \begin{subfigure}[b]{0.49\textwidth}
        \centering
        \includegraphics{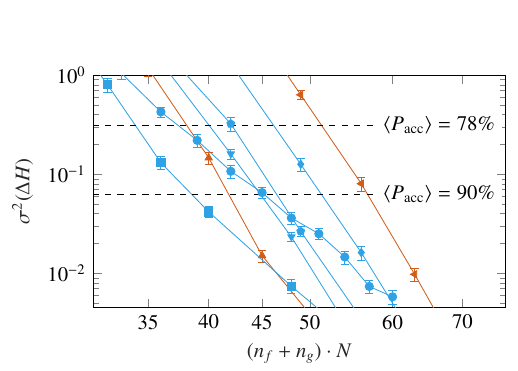}
    \caption{Trajectory length $\tau=2$. The second-order integrators BAB \eqref{eq:BAB} and BABAB \eqref{eq:BABAB} achieve an acceptance probability of $\langle P_{\mathrm{acc}}\rangle < 70\%$ for $(n_f + n_g) \cdot N = 80$ and are thus not included for visualization purposes.}
    \label{fig:Em1_results}
    \end{subfigure}
    \caption{Variance of $\Delta\Hamiltonian$ vs.\ the number of force evaluations per trajectory $(n_f + n_g)\cdot N$ for all non-dominated variants of self-adjoint Hessian-free force-gradient integrators (blue lines) and conventional splitting methods (red lines). For any integrator and step size, 100 trajectories have been computed.}
    \label{fig:Em1_results_both}
\end{figure*}

Starting from a thermalized configuration, we computed $100$ trajectories of length $\tau = 0.1$ with varying step sizes $\stepsize = \tau/N$ ($N \in \mathbb{N}$) for all non-dominated variants of Hessian-free force-gradient integrators and splitting methods of order $\convergenceorder \in \{2,4,6\}$. For these small time steps, the integrators are not affected by numerical instabilities, and the variance of $\Delta \Hamiltonian$ perfectly scales with the expected order $2 \convergenceorder$, as depicted in Figure~\ref{fig:Em1_scaling-phase}.
Moreover, it turns out that in this "scaling phase", the performance of the integrators aligns with the efficiency measure $\mathrm{Eff}_{\mathtt{err}}^{(\convergenceorder)}$. 
However, the depicted region of $\sigma^2(\Delta\Hamiltonian) < 10^{-6}$ is not of particular interest as we are aiming to maximize the step size $\stepsize$ while satisfying \eqref{eq:accuracy_criterion}, i.e., we are interested in the region where $0.01 \leq \sigma^2(\Delta\Hamiltonian) \leq 1$.

Repeating the simulations with $\tau = 2$ yields the results depicted in Figure~\ref{fig:Em1_results}, emphasizing that fourth-order integrators turn out to be the most efficient choice for this setup.
Particularly, the Hessian-free force-gradient integrator ABADABA \eqref{eq:ABADABA} allows for the most efficient computational process, i.e., it provides the best trade-off between accuracy and numerical stability for this particular setup.
Moreover, BADAB \eqref{eq:BADAB} turns out to be an efficient choice as a more stable but less accurate variant. Contrarily, the more accurate but less stable variants turn out to be less efficient.

For fourth-order integrators, the optimal acceptance rate is $\langle P_{\mathrm{acc}}\rangle_{\mathrm{opt}} = \exp(-1/4) \approx 78\%$ \cite{TAKAISHI20006}. Based on the results presented in Figure~\ref{fig:Em1_results}, we can estimate the required number of force evaluations per trajectory \mbox{$(n_f + n_g) \cdot N$} to achieve $\sigma^2(\Delta \Hamiltonian) \approx 0.3121$, resulting in an acceptance rate of approximately 78\%. For the six fourth-order integrators depicted in Figure~\ref{fig:Em1_results}, the computational cost are approximated by taking the two data points closest to the desired acceptance rate, and then performing linear interpolation in logarithmic space. 
The results are summarized in Table~\ref{tab:Em1_cost-comparison}.
In comparison of the most efficient Hessian-free force-gradient integrator ABADABA \eqref{eq:ABADABA} and the splitting method BABABABABAB \eqref{eq:BABABABABAB}, we can conclude that, for this ensemble, the Hessian-free force-gradient integrator requires approximately 89\% of the computational cost to achieve the optimal acceptance rate. 
Unless there are substantial differences in the acceptance probabilities, the choice of the numerical integration scheme will not influence the integrated autocorrelation times, as they are approximating the same system of differential equations numerically. Consequently, we conclude that the framework of Hessian-free force-gradient integrators enables a more efficient computational process compared to conventional splitting methods.

\begin{table}[htb!]
    \centering
    \begin{tabular}{l c}
        \toprule 
        ID & $(n_f + n_g) \cdot N$ for $\langle P_{\mathrm{acc}}\rangle_{\mathrm{opt}}$ \\
        \midrule 
         BADAB & 37.3751\\
         ABADABA & 34.0219 \\
         BABABABABAB & 38.1049 \\
         BADABADAB & 40.0282 \\
         BADABABADAB & 46.1456 \\
         ABADABADABA & 42.0871 \\
         \bottomrule
    \end{tabular}
    \caption{Comparison of the required computational cost $(n_f + n_g) \cdot N$ to achieve the optimal acceptance rate $\langle P_{\mathrm{acc}}\rangle_{\mathrm{opt}} \approx 78\%$ for fourth-order integrators. The results are obtained by performing a linear interpolation in logarithmic space of the two data points in Figure~\ref{fig:Em1_results} that are closest to the optimal acceptance rate.}
    \label{tab:Em1_cost-comparison}
\end{table}

Note that we consider rather large fermion masses here, highlighting that the relative stability threshold $\EffStab$ is already important for larger fermion masses. 
On the one hand, it is anticipated that the relative stability threshold $\EffStab$ will become increasingly important for smaller fermion masses. 
On the other hand, increasing the lattice size does not affect the stability threshold but the accuracy demands, i.e., smaller step sizes must be employed to achieve the same accuracy. Consequently, it is anticipated that the efficiency measure $\EffErr{\convergenceorder}$, which assesses the accuracy at the same computational cost, becomes increasingly important for larger lattices.
This interplay has to be studied further to examine in which situations a certain integrator provides the best possible trade-off.

\subsection{Twisted-mass simulations}\label{sec:tm-simulations}
The linear stability analysis considers the harmonic oscillator as a test equation to determine the stability threshold $z_*$. 
According to the aforementioned hypothesis for interacting field theories~\cite{EDWARDS1997375,instabilityMD}, numerical instabilities in lattice QCD simulations can be likened to that of a collection of oscillator modes, with the onset of the instability caused by the mode of highest frequency $\omega_{\mathtt{max}}$.
We aim for providing numerical evidence on the accuracy of the stability criterion $|\stepsize \omega_{\mathtt{max}}| < z_*$ when performing lattice QCD simulations. 

To investigate the bounds of the stability condition in connection to a lattice QCD setup, we make use of the twisted mass formulation \cite{Frezzotti:2000nk,Frezzotti:2003ni}. Here, the twisted mass term $a \mu$ introduces an UV cut-off, such that $\lambda_{\mathtt{min}}(DD^\dagger+ a^2\mu^2)\ge  a^2\mu^2$. At so-called maximal twist, i.e., the PCAC mass is tuned to zero, the probability density of eigenvalues peaks at the bounds, see \cite{Alexandrou:2016izb}. We investigate the stability bound by using $N_f = 2$ twisted mass fermions with a clover term  $c_{sw} = 1.57551$ and a bare mass parameter of $\kappa = 0.137322$, see \cite{ETM:2015ned}. 
We consider a setup with one Hasenbusch mass \cite{Hasenbusch2001} and a nested integrator \cite{SEXTON1992665,URBACH200687} with three integration levels, where we set the Iwasaki pure gauge action on the level employing the smallest step size (micro level). For the Hasenbusch mass, we select $a\kappa\mu_1 = 0.1$ and set the ratio operator with lower bound $\lambda_{\min} \ge (\mu_1^2 - \mu_0^2)/\mu_0^2\approx \mu_1^2/\mu_0^2$ to the level employing the largest step size (macro level). 
The additional Hasenbusch term $D^\dagger D + \mu_1^2$ is set to the intermediate level.
To ensure that numerical instabilities are not caused by the sub-macro levels, we set the multirate factor of the intermediate level to $M=2$, i.e., the integrator on this level is employed by computing two steps with half the step size.

In a first setup, we use the same conventional splitting method BABAB \eqref{eq:BABAB} for all integration levels. 
On the one hand, we consider the minimum-error variant with $\lambda = \lambda_{\mathtt{err}}$ chosen according to \eqref{eq:OMF2_min-error_coeff} with $\EffErr{2}(\lambda_{\mathtt{err}}) \approx 29.24$ and \mbox{$z_*(\lambda_{\mathtt{err}}) \approx 2.5531$}.
On the other hand, we consider $\lambda = \lambda_{\mathtt{stab}} = 0.25$, resulting in two steps of BAB \eqref{eq:BAB} with step size $\stepsize/2$, which maximizes the relative stability threshold and yields $\EffErr{2}(\lambda_{\mathtt{stab}}) \approx 10.73$ and \mbox{$z_*(\lambda_{\mathtt{stab}}) = 4$}.
When applied to the harmonic oscillator \eqref{eq:harmonic_oscillator}, selecting $\lambda_{\mathtt{stab}}$ instead of $\lambda_{\mathtt{err}}$ allows for the utilization of a \mbox{$z_*(\lambda_{\mathtt{stab}})/z_*(\lambda_{\mathtt{err}}) \approx 1.5667$} times larger step size.

\begin{figure*}[ht!]
    \centering
    \includegraphics{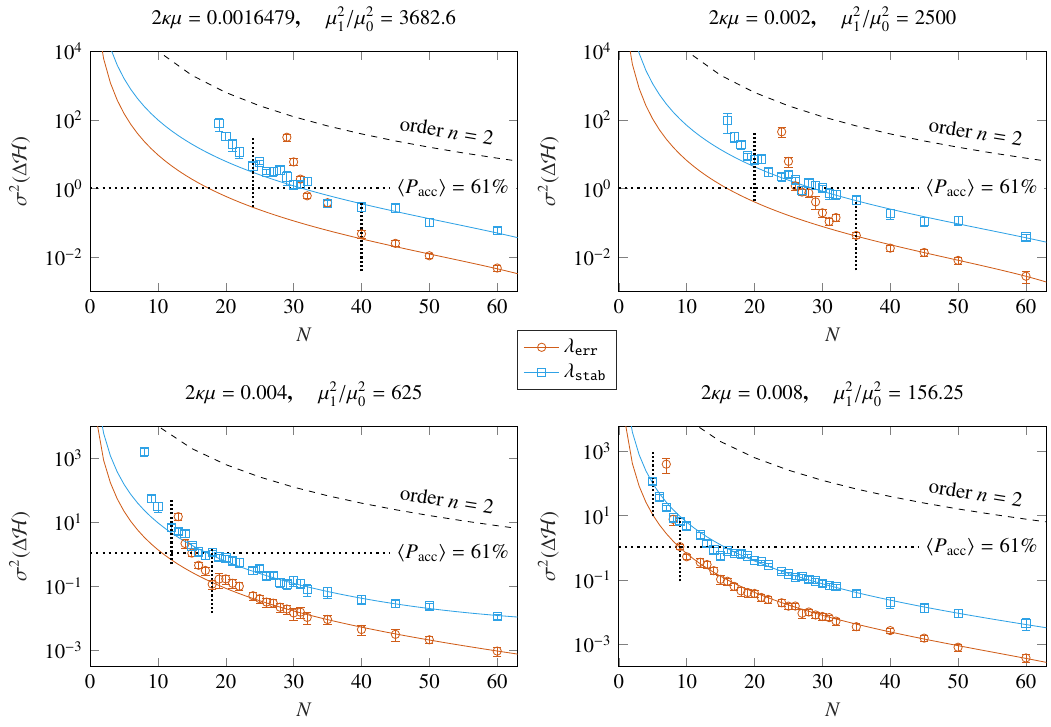}
    \caption{Twisted-mass simulations. Variance of $\Delta\Hamiltonian$ vs.~the number of time steps $N$ per trajectory for the splitting method BABAB \eqref{eq:BABAB} with $\lambda = \lambda_{\mathtt{err}}$ ($\circ$) and $\lambda = \lambda_{\mathtt{stab}}$ ($\square$). The solid lines denote the least squares fitting function $f(N)$. 
    The vertical dotted lines indicate the number of time steps for which the integrator becomes unstable, see \eqref{eq:instability_criterion}. The horizontal dotted lines indicate the optimal acceptance probability of $\langle P_{\mathrm{acc}}\rangle_{\mathrm{opt}} = \exp(-1/2)\approx 61\%$. The dashed lines are reference lines for convergence order $\convergenceorder=2$. For any $N$, $30$ trajectories of length $\tau=1.0$ have been computed.}
    \label{fig:twisted_mass_results}
\end{figure*}

For varying number of time steps $N$ on the macro level, we computed 30 trajectories of length $\tau = 1.0$. The results are depicted in Figure~\ref{fig:twisted_mass_results}. 
Firstly, the numerical results confirm that the minimum-error variant is more accurate than the maximum-stability variant. 
Furthermore, we estimated the number of time steps $N$ per trajectory for which the integrators become unstable. 
As a criterion for instability, we choose
\begin{equation}\label{eq:instability_criterion}
    \sigma^2(\Delta \Hamiltonian(N)) + 1.5 \cdot \mathrm{std}\left( \sigma^2(\Delta \Hamiltonian(N))\right) > f(N),
\end{equation}
where $\sigma^2(\Delta \Hamiltonian(N))$ denotes the variance of $\Delta \Hamiltonian$ when computing $N$ time steps per trajectory, $\mathrm{std}(\sigma^2(\Delta \Hamiltonian(N)))$ its standard deviation, and $f(N) = c_0 + \frac{c_1}{N^{2\convergenceorder}}$ is a least squares fitting function of the data $(N,\sigma^2(\Delta \Hamiltonian(N)))$ for
\begin{equation}\label{eq:N_min}
    N > \begin{cases}
        40, & \lambda = \lambda_{\mathtt{err}}, \\
        26, & \lambda = \lambda_{\mathtt{stab}}.
    \end{cases}
\end{equation}
We denote the onset of instability, i.e., the largest integer satisfying \eqref{eq:instability_criterion}, as $N_{\mathtt{min}}$.
The estimates for $N_{\mathtt{min}}$ are marked in Figure~\ref{fig:twisted_mass_results} by the dotted vertical lines and summarized in Table~\ref{tab:twisted_mass_ratios}. They indicate that the maximum-stability version allows for a larger step size before becoming unstable, as expected. 
Furthermore, Table~\ref{tab:twisted_mass_ratios} contains the ratios between the two integrator variants, emphasizing that the theoretical ratio of $z_*(\lambda_{\mathtt{stab}})/z_*(\lambda_{\mathtt{err}}) \approx 1.5667$ provides a good estimate for evaluating the numerical stability properties of decomposition algorithms when applied to lattice QCD simulations.
Additionally, the results demonstrate that, as the ratio $\mu_1^2/\mu_0^2$ increases, the maximum-stability version becomes increasingly competitive in achieving the optimal acceptance rate of $\langle P_{\mathrm{acc}}\rangle_{\mathrm{opt}} \approx 61\%$ at minimal computational cost. In particular, for $\mu_1^2/\mu_0^2 = 156.25$, the minimum-error variant achieves the optimal acceptance rate for larger step sizes. In contrast, for $\mu_1^2/\mu_0^2 = 3682.6$, the maximum-stability variant allows for fewer time steps per trajectory.

\begin{figure*}[ht!]
    \centering
    \includegraphics{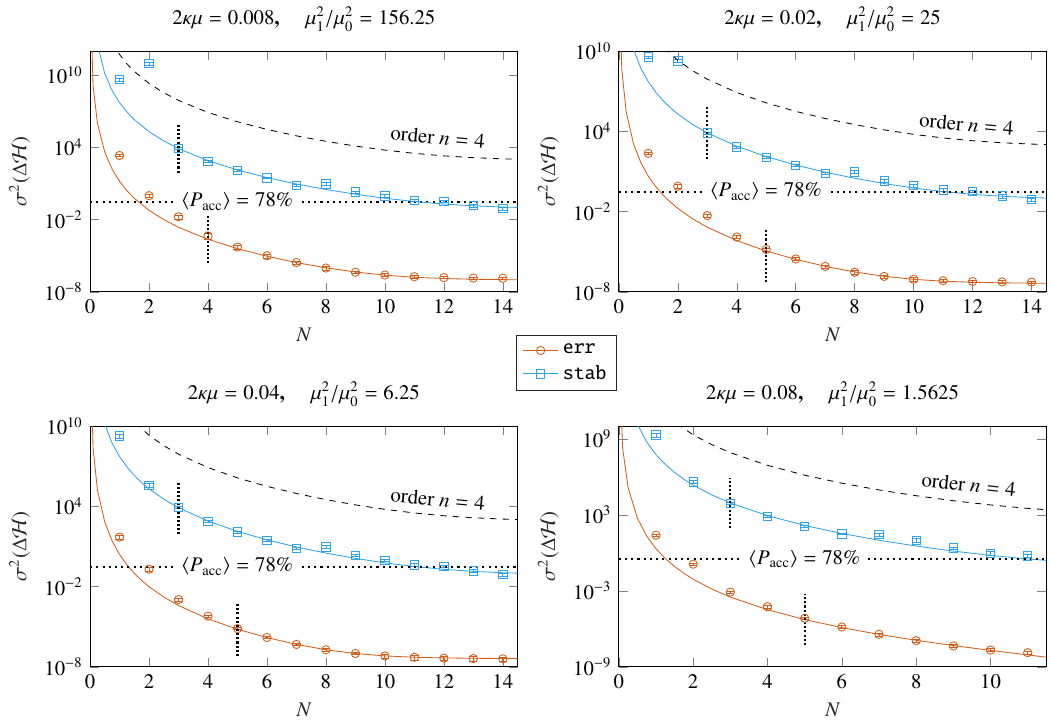}
    \caption{Twisted-mass simulations. Variance of $\Delta\Hamiltonian$ vs.~the number of time steps $N$ per trajectory for the Hessian-free force-gradient integrator ABADABA \eqref{eq:ABADABA} in its minimum-error variant \eqref{eq:ABADABA_min-error} ($\circ$) and its maximum-stability variant \eqref{eq:ABADABA_max-stab} ($\square$).
    The solid lines denote the least squares fitting function $f(N)$. 
    The vertical dotted lines indicate the number of time steps for which the integrator becomes unstable, see \eqref{eq:instability_criterion}.
    The horizontal dotted lines indicate the optimal acceptance probability of $\langle P_{\mathrm{acc}}\rangle_{\mathrm{opt}} = \exp(-1/4) \approx 78\%$.
    The dashed lines are reference lines for convergence order $\convergenceorder=4$. For any $N$, $30$ trajectories of length $\tau=0.35$ have been computed.}
    \label{fig:twisted_mass_ABADABA}
\end{figure*}

In a second setup, we consider variants of the Hessian-free force-gradient integrator ABADABA \eqref{eq:ABADABA} for all integration levels. In particular, we compare the minimum-error variant \eqref{eq:ABADABA_min-error} with the variant maximizing the stability threshold that is obtained by choosing $a_1$ and $c_2$ according to \eqref{eq:ABAXABA_coeffs} with
\begin{equation}\label{eq:ABADABA_max-stab}
    b_1 = 0.050446855826563.
\end{equation}
Maximizing the stability threshold yields $z_* \approx 4.6703$ and thus a larger relative stability threshold $\EffStab \approx 1.1676$. However, this variant is way less accurate with $\EffErr{4} \approx 0.24$. 
According to the linear stability analysis, the maximum-stability variant allows for the utilization of a $$z_*(\mathtt{stab}) / z_*(\mathtt{err}) = 4.6703/3.1377 \approx 1.4884$$ times larger step size compared to the minimum-error variant.
For varying number of time steps $N$ on the macro level, we computed $30$ trajectories of length $\tau = 0.35$. The results are depicted in Figure~\ref{fig:twisted_mass_ABADABA}. 
Similar to the investigations of the splitting method BABAB in Figure~\ref{fig:twisted_mass_results}, we determined the onset of instability via \eqref{eq:instability_criterion}, where for the determination of the least squares fitting functions, we use for both integrator variants the data $(N, \sigma^2(\Delta \Hamiltonian(N)))$ for $N>6$.
These estimates are indicated in Figure~\ref{fig:twisted_mass_ABADABA} by the dotted vertical lines. The estimates, along with the ratios between the two integrator variants, are summarized in Table~\ref{tab:twisted_mass_ABADABA}.
We once again observe that the theoretical ratio of $z_*(\mathtt{stab})/z_*(\mathtt{err}) \approx 1.4884$ provides a good estimate for evaluating the numerical stability.
However, for the ratios $\mu_1^2/\mu_0^2 \in \{156.25,25\}$, we observe that the maximum-stability variant exhibits a significant increase of $\sigma^2(\Delta \Hamiltonian)$ after passing the onset of instability. This can be attributed to the significant inaccuracy of the maximum-stability version. The instabilities commence at extremely high errors, where none of the samples have been accepted, and we have already exited the region of interest.
In summary, across all four scenarios, the minimum-error variant consistently achieves the optimal acceptance probability of $\langle P_{\mathrm{acc}}\rangle_{\mathrm{opt}} \approx 78\%$ while requiring less time steps per trajectory and thus is the more efficient choice. 
This once more emphasizes that maximizing the stability threshold without considering the accuracy is not advisable, as the accuracy becomes the bottleneck. 
In contrast, the minimum-error variant possesses a marginally smaller stability threshold while simultaneously achieving significantly enhanced accuracy.  

\begin{table}[ht!]
    \centering
    \begin{tabular}{@{} r c c c @{}}
        \toprule
        $\mu_1^2 / \mu_0^2$ & $N_{\mathtt{min}}(\lambda_{\mathtt{err}})$ & $N_{\mathtt{min}}(\lambda_{\mathtt{stab}})$ & $N_{\mathtt{min}}(\lambda_{\mathtt{err}}) / N_{\mathtt{min}}(\lambda_{\mathtt{stab}})$   \\
        \midrule 
        3682.60 & 40 & 24 & 1.6667 \\
        2500.00 & 35 & 20 & 1.7500 \\
        625.00 & 18 & 12 & 1.5000 \\
        156.25 & 9 & 5 & 1.8000\\
        \bottomrule
    \end{tabular}%
    \caption{Twisted-mass simulations. Estimates of the onset of instability $N_{\mathtt{min}}$ according to \eqref{eq:instability_criterion} for the splitting method BABAB \eqref{eq:BABAB} with $\lambda \in \{ \lambda_{\mathtt{err}},\lambda_{\mathtt{stab}}\}$ and different values of $\mu_1^2 / \mu_0^2$. The last column contains the ratios between the two integrator variants, indicating how much larger step sizes can be employed for the maximum-stability variant compared to the minimum-error variant.}
    \label{tab:twisted_mass_ratios}
\end{table}

\begin{table}[htb!]
    \centering
    \begin{tabular}{@{} l c c c @{}}
        \toprule
        $\mu_1^2 / \mu_0^2$ & $N_{\mathtt{min}}(\mathtt{err})$  & $N_{\mathtt{min}}(\mathtt{stab})$ & $N_{\mathtt{min}}(\mathtt{err}) / N_{\mathtt{min}}(\mathtt{stab}) $   \\
        \midrule 
        156.25 & 4 & 3 & 1.3333 \\
        \phantom{1}25.00 & 5 & 3 & 1.6667 \\
        \phantom{15}6.25 & 5 & 3 & 1.6667 \\
        \phantom{15}1.5625 & 5 & 3 & 1.6667\\
        \bottomrule
    \end{tabular}%
    \caption{Twisted-mass simulations. Estimates of the onset of instability $N_{\mathtt{min}}$ according to \eqref{eq:instability_criterion} for the Hessian-free force-gradient integrator ABADABA \eqref{eq:ABADABA} in its minimum-error variant \eqref{eq:ABADABA_min-error} and its maximum-stability variant \eqref{eq:ABADABA_max-stab} for different values of $\mu_1^2 / \mu_0^2$. The last column contains the ratios between the two integrator variants, indicating how much larger step sizes can be employed for the maximum-stability variant compared to the minimum-error variant.}
    \label{tab:twisted_mass_ABADABA}
\end{table}

\section{Conclusion and outlook}
In this work, we conducted a linear stability analysis for force-gradient integrators and their Hessian-free counterparts. 
When applied to linear separable Hamiltonian systems, conventional force-gradient integrators and Hessian-free force-gradient integrators are equivalent, i.e., their linear stability analysis coincides.
For the family of self-adjoint integrators with $s \leq 6$ stages, we performed detailed stability investigations, augmented by the accuracy investigations performed in \cite{omelyan2003symplectic,schäfers2024hessianfree}. 
The investigations show that the efficiency measure $\EffErr{\convergenceorder}$ is way more sensitive to alterations in the integrator coefficients than the relative stability threshold $\EffStab$. As a consequence, the family of minimum-error methods derived in \cite{omelyan2003symplectic,schäfers2024hessianfree} contains promising variants. 
Incorporating the relative stability threshold $\EffStab$ allows to detect minimum-error variants that show desirable stability properties. Particularly, for convergence order $\convergenceorder = 4$, conventional force-gradient integrators and Hessian-free force-gradient integrators turn out to be promising choices.
Numerical tests demonstrate that Hessian-free force-gradient integrators offer a more efficient computational process compared to conventional splitting methods. 
Although the efficiency measure $\EffErr{\convergenceorder}$ provided insights into the accuracy of the integrators, it was insufficient to explain the superior efficiency of certain variants. 
The relative stability threshold $\EffStab$ reveals that the efficiency of the integrators is strongly influenced by their numerical stability.
Furthermore, the simulations for the two-dimensional Schwinger model provide numerical evidence indicating no significant difference in the numerical stability of force-gradient integrators and their Hessian-free counterparts.
Lattice QCD simulations with two heavy Wilson fermions reveal that, already for large fermion masses, stability-enhanced variants turn out to be more efficient than more accurate but less stable variants. 
In the numerical tests, a Hessian-free force-gradient integrator, \mbox{ABADABA} \eqref{eq:ABADABA}, achieves the same acceptance probability as the most efficient conventional splitting method, BABABABABAB \eqref{eq:BABABABABAB}, while requiring less than $90\%$ of the computational cost. 
Further numerical tests of the accuracy of the stability threshold by investigating $N_f = 2$ twisted mass fermions and nested integrators highlight that the linear stability threshold derived from investigating the harmonic oscillator emerges as a reliable stability criterion for realistic lattice QCD simulations.
When increasing the lattice volume, the accuracy is expected to become more important as the numerical stability does not depend on the volume. On the other hand, for smaller fermion masses, the numerical stability becomes more important so that preference should be given to integrators with a higher value for $\EffStab$. Depending on the particular ensemble, it has to be checked which variant provides the best possible trade-off between accuracy and numerical stability for this particular problem.

In general, the efficiency of the integrators can be further increased by the use of multiple time stepping techniques \cite{SEXTON1992665,URBACH200687}. 
For this purpose, we aim to provide a general framework for these \emph{nested integrators}, including an order theory, a refined error analysis, a linear stability analysis, as well as generalizations to the efficiency measures $\EffErr{\convergenceorder}$ and $\EffStab$. 
Moreover, a next step is to enhance the efficiency measure $\EffErr{\convergenceorder}$. Taking the standard Euclidean norm for the efficiency measure $\EffErr{\convergenceorder}$ assumes that all error terms contribute equally to the overall error. In practice, this is not necessarily the case. Deriving problem-dependent weights to obtain a weighted norm and to minimize this norm instead will provide a more appropriate measure for the computational efficiency when applied to a specific ensemble. 

\section*{CRediT authorship contribution statement}
\textbf{Kevin Schäfers:} 
Writing - original draft, Validation, Software, Methodology, Investigation, Conceptualization.
\textbf{Jacob Finkenrath:} Writing - original draft, Validation, Software, Investigation.
\textbf{Michael Günther:} Writing - review \& editing, Supervision, Funding acquisition.
\textbf{Francesco Knechtli:} Writing- review \& editing, Supervision, Project administration, Funding acquisition.

\section*{Declaration of competing interest}
The authors declare that they have no known competing financial interests or personal relationships that could have appeared to influence the work reported in this paper.

\section*{Data availability}
An extended version of openQCD v2.4 (\href{https://luscher.web.cern.ch/luscher/openQCD/}{https://\allowbreak{}luscher.\allowbreak{}web.\allowbreak{}cern.\allowbreak{}ch/\allowbreak{}luscher/\allowbreak{}openQCD/}) including the class of Hessian-free force-gradient integrators is opensource and available at the GitHub repository: \newline \href{https://github.com/KevinSchaefers/openQCD_force-gradient}{https://github.com/Kevin\allowbreak{} Schaefers/\-openQCD\_force-gradient}.

\section*{Acknowledgements}
This work is supported by the German Research Foundation (DFG) research unit FOR5269
"Future methods for studying confined gluons in QCD".
The simulations from Section~\ref{sec:Em1} were carried out on the PLEIADES cluster at the University of Wuppertal, which was supported by the DFG and the Federal Ministry of Education and Research (BMBF), Germany.
The authors acknowledge fruitful discussions with Stefan Schaefer.

J.~Finkenrath received financial support from the Inno4scale project, which received funding from the European High-Perfor\-mance Computing Joint Undertaking (JU) under Grant Agreement No.\ 101118139. The JU receives support from the European Union’s Horizon Europe Programme. 
J.~Finkenrath acknowledges financial support by the Eric \& Wendy Schmidt Fund for Strategic Innovation through the CERN Next Generation Triggers project under grant agreement number SIF-2023-004.

Funded by the European Union. Views and opinions expressed are however those of the author(s) only and do not necessarily reflect those of the European Union or the European Research Council Executive Agency (ERCEA). Neither the European Union nor the ERCEA can be held responsible for them.

\appendix 
\section{Decomposition algorithms for lattice QCD simulations}\label{app:LieGroup}
\sloppy For gauge field simulations in lattice QCD on a four-dimensional lattice of size $V = T \times L^3$ with lattice spacing $a$, the Hamiltonian considered in the HMC algorithm reads
$$ \Hamiltonian([\U],[\MatrixP]) = \Kinetic([\MatrixP]) + \Action([\U]),$$
with kinetic energy $\Kinetic([\MatrixP]) = \tfrac{1}{2} \sum_{x,\mu} \mathrm{tr}(\MatrixP_{x,\mu}^2)$ and action $\Action([\U])$. Here, the links $\U_{x,\mu}$, connecting the sizes $x$ and $x + a \hat{\mu}$, denote matrix representations of elements of $\SUthree$, whereas $\MatrixP_{x,\mu}$ are matrix representations of the momenta in the Lie algebra $\suthree$ of traceless and anti-Hermitian matrices.
Any element $\MatrixP \in \suthree$ can be expressed as $\MatrixP = p^i \Generator_i$ where $\Generator_i$ denote the generators of the Lie algebra. The linear differential operators $\diffoperator_i$ act on the Lie group elements $\U$ as $\diffoperator_i \U = -\Generator_i \U$. They are gauge-covariant generalizations of the vector field $\partial / \partial q^i$.  
As the Lie group $\SUthree$ is a semi-simple matrix Lie group, the vector fields of the kinetic energy and the action are given by~\cite{kennedy2013shadow}
\begin{align}\label{eq:Hamiltonian_vectorfields_SU3}
    \Kineticvf &= p^i \diffoperator_i, & \Actionvf &= - \diffoperator_i(\Action) \frac{\partial}{\partial p_i}. 
\end{align}
Consequently, the equations of motion read 
\begin{equation}\label{eq:equations_of_motion_SU3}
\begin{aligned}
    \dot{\U} &= p^i \diffoperator_i(\U) = -p^i \Generator_i \U = -\MatrixP\U, & \U(0) &= \U_0, \\
    \dot{\MatrixP} &= -\diffoperator_i(\Action) \frac{\partial \MatrixP}{\partial p_i} = - \diffoperator_i(\Action) \Generator^i, & \MatrixP(0) &= \MatrixP_0. 
\end{aligned}
\end{equation}
As for Euclidean space, the splitting $\Hamiltonian = \Kinetic + \Action$ yields two easily solvable Hamiltonian systems 
\begin{align*}
    \begin{pmatrix}
         \dot{\U} \\ \dot{\MatrixP}
    \end{pmatrix} &= \begin{pmatrix}
        -\MatrixP\U \\ \0
    \end{pmatrix}, & \begin{pmatrix}
         \dot{\U} \\ \dot{\MatrixP}
    \end{pmatrix} &= \begin{pmatrix}
         \0 \\ -\diffoperator_i(\Action) \Generator^i 
    \end{pmatrix}.
\end{align*}
The formal solution of these two subsystems can be expressed in terms of the vector fields \eqref{eq:Hamiltonian_vectorfields_SU3} as 
\begin{subequations}\label{eq:flows_subsystems_SU3}
\begin{align}
    &\e^{t \Kineticvf} \begin{pmatrix}
        \U_0 \\ \MatrixP_0
    \end{pmatrix} = \begin{pmatrix}
        \exp(-t\MatrixP_0)\U_0 \\ \MatrixP_0
    \end{pmatrix}, \label{eq:link_update_SU3} \\
    &\e^{t \Actionvf} \begin{pmatrix}
        \U_0 \\ \MatrixP_0
    \end{pmatrix} = \begin{pmatrix}
         \U_0 \\ \MatrixP_0 - t \diffoperator_i(\Action)(\U_0)\Generator^i
    \end{pmatrix},\label{eq:momentum_update_SU3}
\end{align}
\end{subequations}
with $\exp$ denoting the matrix exponential.
We refer to \eqref{eq:link_update_SU3} as a link update and to \eqref{eq:momentum_update_SU3} as a momentum update.\medskip

\noindent\textbf{Splitting methods.}
Splitting methods compute a numerical approximation to the exact flow $\e^{\stepsize(\Kineticvf + \Actionvf)}$ by composing evaluations of the flows \eqref{eq:flows_subsystems_SU3}. This leads to numerical integration schemes of the form 
$$ \Phi_\stepsize = \e^{a_1 \stepsize \Kineticvf} \e^{b_1 \stepsize \Actionvf} \cdots \e^{a_s \stepsize \Kineticvf} \e^{b_s \stepsize \Actionvf}.$$

\noindent\textbf{Force-gradient integrators.}
For the Hamiltonian system \eqref{eq:equations_of_motion_SU3}, the force-gradient term reads 
$$ \ForceGradientvf = [\Actionvf , [\Kineticvf, \Actionvf]] = 2 \diffoperator^j(\Action) \diffoperator_j \diffoperator_i(\Action) \frac{\partial}{\partial p_i}.$$
By defining force-gradient steps 
\begin{equation}\label{eq:fg_update_SU3}
\begin{aligned}
    \e^{b_{\ell} \stepsize \Actionvf + c_{\ell} \stepsize^3 \ForceGradientvf} \begin{pmatrix}\U_0 \\ \MatrixP_0 \end{pmatrix} = &\begin{pmatrix}
        \U_0 \\ \MatrixP_0 
    \end{pmatrix} + \begin{pmatrix}
         \0 \\ -b_{\ell} \stepsize \diffoperator_i(\Action)(\U_0) \Generator^i
    \end{pmatrix} \\
    &+ \begin{pmatrix}
        \0 \\ 2c_{\ell} \stepsize^3 \diffoperator^j(\Action)(\U_0) \diffoperator_j \diffoperator_i(\Action)(\U_0) \Generator^i  
    \end{pmatrix},
\end{aligned}
\end{equation}
a force-gradient integrator computes a numerical approximation to the exact flow $\e^{\stepsize(\Kineticvf + \Actionvf)}$ by a composition of link updates \eqref{eq:link_update_SU3} and force-gradient updates \eqref{eq:fg_update_SU3}, resulting in integrators of the form 
$$ \Phi_\stepsize = \e^{a_1 \stepsize \Kineticvf} \e^{b_1 \stepsize \Actionvf + c_1 \stepsize^3 \ForceGradientvf} \cdots \e^{a_s \stepsize \Kineticvf} \e^{b_s \stepsize \Actionvf + c_s \stepsize^3 \ForceGradientvf}.$$

\noindent\textbf{Hessian-free force-gradient integrators.}
In the Hessian-free framework, the force-gradient steps \eqref{eq:fg_update_SU3} are approximated via 
\begin{equation} \label{eq:approx_fg-step_SU3}
\begin{aligned}
    \e^{b_{\ell} \stepsize \hat{\mathcal{D}}(\stepsize,b_\ell,c_\ell) }&\begin{pmatrix} \U_0 \\ \MatrixP_0\end{pmatrix} \\
    \coloneqq &\begin{pmatrix} \U_0 \\ \MatrixP_0-b_\ell \stepsize \diffoperator_i(\Action) \left( \exp\left( - \frac{2 c_\ell \stepsize^2}{b_\ell} \boldsymbol{F}^j \Generator_j \right) \U_0 \right) \Generator^i\end{pmatrix},
\end{aligned}
\end{equation}
where $\boldsymbol{F}^j \diffoperator_j = \diffoperator^j(\Action)(\U_0) \diffoperator_j$ is regarded as a frozen vector field, i.e., $\diffoperator_j$ acting on $\boldsymbol{F}^j$ is defined to be zero.
This approximation introduces an error of order $\mathcal{O}(h^5)$ \cite{schäfers2024hessianfree}. Replacing the force-gradient steps \eqref{eq:fg_update_SU3} by the approximated ones \eqref{eq:approx_fg-step_SU3} results in a Hessian-free force-gradient integrator 
$$ \Phi_\stepsize = \e^{a_1 \stepsize \Kineticvf} \e^{b_1 \stepsize \hat{\mathcal{D}}(\stepsize,b_1,c_1)} \cdots \e^{a_s \stepsize \Kineticvf} \e^{b_s \stepsize \hat{\mathcal{D}}(\stepsize,b_s,c_s)}.$$

\section{Detailed overview of minimum-error variants}\label{app:Integrators}
Throughout the paper, certain variants of splitting methods \cite{mclachlan2002splitting}, force-gradient integrators \cite{omelyan2003symplectic}, and Hessian-free force-gradient integrators \cite{schäfers2024hessianfree} have been detected as non-dominated variants as there is no other variant that simultaneously performs better in terms of both accuracy (measured by the efficiency measure $\EffErr{\convergenceorder}$) and numerical stability (measured by the relative stability threshold $\EffStab$). 
For completeness, we provide the definitions and integrator coefficients of the minimum-error variants of all decomposition algorithms that have been declared as non-dominated and have not been defined already throughout the paper. The integrator coefficients are taken from \cite{omelyan2003symplectic} and \cite{schäfers2024hessianfree}, respectively.
Moreover, we provide values for both efficiency measures $\EffErr{\convergenceorder}$ and $\EffStab$ for the families of self-adjoint conventional and Hessian-free force-gradient integrators with $s \leq 6$ stages in their respective minimum-error variant, extending the tables \cite[Table~2]{omelyan2003symplectic} and \cite[Table~1]{schäfers2024hessianfree}, respectively.

\subsection{Splitting Methods}

\noindent\textbf{BABABABAB.} 
\begin{subequations}\label{eq:BABABABAB}
    \begin{equation}
    \begin{aligned}
        \Phi_\stepsize = &\e^{b_1 \stepsize \Potentialvf} \e^{a_2 \stepsize \Kineticvf} \e^{b_2 \stepsize \Potentialvf} \e^{(0.5 - a_2) \stepsize \Kineticvf} \e^{(1-2(b_1+b_2)) \stepsize \Potentialvf} \\
        &\quad\e^{(0.5 - a_2) \stepsize \Kineticvf} \e^{b_2 \stepsize \Potentialvf} \e^{a_2 \stepsize \Kineticvf} \e^{b_1 \stepsize \Potentialvf},
    \end{aligned}
    \end{equation}
    with integrator coefficients
    \begin{equation}
    \begin{aligned}
        a_2 &= 0.520943339103990, \\
        b_1 &= 0.164498651557576, \\
        b_2 &= 1.235692651138917. 
    \end{aligned}
    \end{equation}
\end{subequations}

\noindent\textbf{ABABABABA.} 
\begin{subequations}\label{eq:ABABABABA}
    \begin{equation}
    \begin{aligned}
         \Phi_\stepsize = &\e^{a_1 \stepsize \Kineticvf} \e^{b_1 \stepsize \Potentialvf} \e^{a_2 \stepsize \Kineticvf} \e^{(0.5 - b_1) \stepsize \Potentialvf} \e^{(1-2(a_1+a_2)) \stepsize \Kineticvf} \\
         &\quad \e^{(0.5 - b_1) \stepsize \Potentialvf} \e^{a_2 \stepsize \Kineticvf} \e^{b_1 \stepsize \Potentialvf} \e^{a_1 \stepsize \Kineticvf},
    \end{aligned}
    \end{equation}
    with integrator coefficients
    \begin{align}
    \begin{split}
        a_1 &= \phantom{-}0.178617895844809, \\
        a_2 &= -0.066264582669818, \\
        b_1 &= \phantom{-}0.712341831062606.
    \end{split}
    \end{align}
\end{subequations}

\noindent\textbf{BABABABABAB.} 
\begin{subequations}\label{eq:BABABABABAB}
    \begin{equation}
    \begin{aligned}
        \Phi_\stepsize = &\e^{b_1 \stepsize \Potentialvf} \e^{a_2 \stepsize \Kineticvf} \e^{b_2 \stepsize \Potentialvf} \e^{a_3 \stepsize \Kineticvf} \e^{(0.5 - (b_1+b_2)) \stepsize \Potentialvf} \\
        &\quad\e^{(1- 2(a_2+a_3)) \stepsize \Kineticvf} \e^{(0.5 - (b_1+b_2)) \stepsize \Potentialvf} \e^{a_3 \stepsize \Kineticvf} \e^{b_2 \stepsize \Potentialvf} \\
        &\qquad\e^{a_2 \stepsize \Kineticvf} \e^{b_1 \stepsize \Potentialvf},
    \end{aligned}
    \end{equation}
    with integrator coefficients
    \begin{align}
        \begin{split}
            a_2 &= \phantom{-}0.253978510841060, \\
            a_3 &= -0.032302867652700, \\
            b_1 &= \phantom{-}0.083983152628767, \\
            b_2 &= \phantom{-}0.682236533571909.
        \end{split}
    \end{align}
\end{subequations}\medskip

In Table~\ref{tab:splitting_method_overview}, the minimum-error variants of splitting methods with $s \leq 6$ stages, as well as the sixth-order variants with $s=8$ stages, are summarized with their values for both $\EffErr{\convergenceorder}$ and $\EffStab$. For the integrator coefficients of the respective variants, we refer to \cite{omelyan2003symplectic}.

\renewcommand{\arraystretch}{1.2}
\begin{table*}[ht!]
    \centering
    \begin{tabular}{@{}c | l c c l r c c c c@{}}
        \toprule 
        & Integrator & $n_f$ & $n_g$ & $\mathrm{Err}_{\convergenceorder+1}$ & $\EffErr{\convergenceorder}$ & $z_*$ & $\EffStab$ & Remarks & Eqs. \\
        \midrule  
        \multirow{4}{*}{\rotatebox{90}{$\convergenceorder = 2$}} 
        & BAB & 1 & 0 & 0.0932 & 10.73 & 2.0000 & 2.0000 & \cite{verlet1967computer,strang1968construction} & \eqref{eq:BAB} \\
        & ABA & 1 & 0 & 0.0932 & 10.73 & 2.0000 & 2.0000 & \cite{verlet1967computer,strang1968construction} & \eqref{eq:ABA} \\
        & BABAB & 2 & 0 & 0.00855 & 29.24 & 2.5531 & 1.2766 & \cite{mclachlan1995,omelyan2003symplectic} & \eqref{eq:BABAB} \\
        & ABABA & 2 & 0 & 0.00855 & 29.24 & 2.5531 & 1.2766 & \cite{mclachlan1995,omelyan2003symplectic} & \eqref{eq:ABABA} \\
        \midrule
        \multirow{6}{*}{\rotatebox{90}{$\convergenceorder = 4$}} 
        & BABABAB & 3 & 0 & 0.0383 & 0.32 & 1.5734 & 0.5245 & \cite{yoshida1990construction,forest1990fourth} & \\
        & ABABABA & 3 & 0 & 0.0283 & 0.44 & 1.5734 & 0.5245 & \cite{yoshida1990construction,forest1990fourth} & \\
        & BABABABAB & 4 & 0 & 0.000654 & 5.97 & 3.4696 & 0.8674 & \cite{omelyan2003symplectic} & \eqref{eq:BABABABAB} \\
        & ABABABABA & 4 & 0 & 0.000610 & 6.40 & 2.9894 & 0.7474 & \cite{omelyan2003symplectic} & \eqref{eq:ABABABABA} \\
        & BABABABABAB & 5 & 0 & 0.0000270 & 59.26 & 3.1421 & 0.6284 & \cite{omelyan2003symplectic} & \eqref{eq:BABABABABAB} \\
        & ABABABABABA & 5 & 0 & 0.0000518 & 30.91 & 2.9763 & 0.5953 & \cite{omelyan2003symplectic} & \\
        \midrule
        \multirow{2}{*}{\rotatebox{90}{$\convergenceorder = 6$}} 
        & BABABABABABABAB & 7 & 0 & 0.00000609 & 1.40 & 3.1603 & 0.4515 & \cite{omelyan2003symplectic} & \eqref{eq:BABABABABABABAB} \\
        & ABABABABABABABA & 7 & 0 & 0.0000109 & 0.78 & 3.0674 & 0.4382 & \cite{omelyan2003symplectic} & \\
        \bottomrule
    \end{tabular}
    \caption{Collection of splitting methods in their respective minimum-error variants. For all integrators, the number of force evaluations $n_f$ and force-gradient evaluations $n_g$ per time step, the leading error term $\mathrm{Err}_{\convergenceorder + 1}$, the efficiency measure $\EffErr{\convergenceorder}$, the stability threshold $z_*$, the relative stability threshold $\EffStab$, as well as a remark on the earlier appearance of the integrations, are stated. If available, the final column refers to the equations where the respective variant is defined.}
    \label{tab:splitting_method_overview}
\end{table*}

\subsection{Force-gradient integrators}
\noindent\textbf{CABACABAC.} 
\begin{subequations}\label{eq:CABACABAC}
    \begin{equation}
    \begin{aligned}
        \Phi_\stepsize = &\e^{b_1 \stepsize \Potentialvf + c_1 \stepsize^3 \ForceGradientvf} \e^{a_2 \stepsize \Kineticvf} \e^{b_2 \stepsize \Potentialvf} \e^{(0.5 - a_2) \stepsize \Kineticvf} \\
        &\quad\e^{(1-2(b_1+b_2)) \stepsize \Potentialvf + c_3 \stepsize^3 \ForceGradientvf} \e^{(0.5 - a_2) \stepsize \Kineticvf} \e^{b_2 \stepsize \Potentialvf} \\
        &\qquad\e^{a_2 \stepsize \Kineticvf} \e^{b_1 \stepsize \Potentialvf + c_1 \stepsize^3 \ForceGradientvf},
    \end{aligned}
    \end{equation}
    with integrator coefficients
    \begin{align}
        \begin{split}
            a_2 &=  0.1921125277429464, \\
            b_1 &= 0.0585187261345562, \\
            b_2 &= 0.2852162240687091, \\
            c_1 &= 0.0004339598806816, \\
            c_3 &= 0.0024274752596631. 
        \end{split}
    \end{align}
\end{subequations}

\noindent\textbf{ABACABACABA.} 
\begin{subequations}\label{eq:ABACABACABA}
    \begin{equation}
    \begin{aligned}
        \Phi_\stepsize = &\e^{a_1 \stepsize \Kineticvf} \e^{b_1 \stepsize \Potentialvf} \e^{a_2 \stepsize \Kineticvf} \e^{b_2 \stepsize \Potentialvf + c_2 \stepsize^3 \ForceGradientvf} \e^{(0.5-(a_1+a_2)) \stepsize \Kineticvf} \\
        &\quad\e^{(1-2(b_1+b_2)) \stepsize \Potentialvf}\e^{(0.5-(a_1+a_2)) \stepsize \Kineticvf} \e^{b_2 \stepsize \Potentialvf + c_2 \stepsize^3 \ForceGradientvf} \e^{a_2 \stepsize \Kineticvf} \\
        &\qquad \e^{b_1 \stepsize \Potentialvf} \e^{a_1 \stepsize \Kineticvf},
    \end{aligned}
    \end{equation}
    with integrator coefficients
    \begin{align}
    \begin{split}
        a_1 &= 0.0641910886681624, \\
        a_2 &= 0.1919807940455741, \\
        b_1 &= 0.1518179640276466, \\
        b_2 &= 0.2158369476787619, \\
        c_2 &= 0.0009628905212025. 
    \end{split}
    \end{align}
\end{subequations}

The family of self-adjoint force-gradient integrators with \mbox{$s \leq 6$} stages has been investigated in \cite{omelyan2003symplectic} by deriving the minimum-error variants based on the efficiency measure \eqref{eq:EffErr} and assuming $\xi = 2$. In Table~\ref{tab:force-gradient_overview}, the classification for the respective variants has been augmented by the stability threshold $z_*$ and the relative stability threshold $\EffStab$.

\subsection{Hessian-free force-gradient integrators}
\noindent\textbf{ABADABA.}
The minimum-error variant of the Hessian-free force-gradient integrator ABADABA \eqref{eq:ABADABA} is obtained by choosing $a_1$ and $c_2$ according to \eqref{eq:ABAXABA_coeffs} with
\begin{equation}\label{eq:ABADABA_min-error}
    b_1 = 0.247597680043986.
\end{equation}

\noindent\textbf{BADABADAB.}
The minimum-error variant of the Hessian-free force-gradient integrator BADABADAB \eqref{eq:BADABADAB} is obtained by choosing $b_1$ and $c_2$ according to \eqref{eq:BAXABAXAB_coeffs} with 
\begin{equation}
    \begin{aligned}\label{eq:BADABADAB_min-error}
        a_2 &= 0.219039425103133, \\
        b_2 &= 0.311000565033563.
    \end{aligned}
\end{equation}

\noindent\textbf{BADABABADAB.}
\begin{subequations}\label{eq:BADABABADAB}
    \begin{align}
    \begin{split}
        \Phi_\stepsize = &\e^{b_1 \stepsize \Potentialvf} \e^{a_2 \stepsize \Kineticvf} \e^{b_2 \stepsize \mathcal{\hat{D}}(\stepsize,b_2,c_2)} \e^{a_3 \stepsize \Kineticvf} \e^{(0.5-(b_1+b_2)) \stepsize \Potentialvf}\\
        &\quad \e^{(1-2(a_2+a_3))\stepsize \Kineticvf} \e^{(0.5-(b_1+b_2)) \stepsize \Potentialvf} \e^{a_3 \stepsize \Kineticvf} \\
        &\qquad \e^{b_2 \stepsize \mathcal{\hat{D}}(\stepsize,b_2,c_2)} \e^{a_2 \stepsize \Kineticvf} \e^{b_1 \stepsize \Potentialvf},
    \end{split}
    \end{align}
    with integrator coefficients
    \begin{align}
    \begin{split}
        a_2 &= 0.201110227930330, \\ 
        a_3 &= 0.200577842713366, \\
        b_1 &= 0.065692416344302, \\
        b_2 &= 0.264163604920340, \\
        c_2 &= 0.001036943019757. 
    \end{split}
    \end{align}
\end{subequations}

\noindent\textbf{ABADABADABA.}
\begin{subequations}\label{eq:ABADABADABA}
    \begin{align}
    \begin{split}
        \Phi_\stepsize = &\e^{a_1 \stepsize \Kineticvf} \e^{b_1 \stepsize \Potentialvf} \e^{a_2 \stepsize \Kineticvf} \e^{b_2 \stepsize \mathcal{\hat{D}}(\stepsize,b_2,c_2)} \e^{(0.5-(a_1+a_2)) \stepsize \Kineticvf} \\
        &\quad \e^{(1-2(b_1+b_2)) \stepsize \Potentialvf} \e^{(0.5-(a_1+a_2))\stepsize \Kineticvf} \e^{b_2 \stepsize \mathcal{\hat{D}}(\stepsize,b_2,c_2)} \\
        &\qquad\e^{a_2 \stepsize \Kineticvf} \e^{b_1 \stepsize \Potentialvf} \e^{a_1 \stepsize \Kineticvf},
    \end{split}
    \end{align}
    with integrator coefficients
    \begin{align}
    \begin{split}
        a_1 &= 0.062702644098210, \\
        a_2 &= 0.193174566017780, \\
        b_1 &= 0.149293739165427, \\
        b_2 &= 0.220105234408407, \\
        c_2 &= 0.000966194415594.
    \end{split}
    \end{align}
\end{subequations}

The family of self-adjoint Hessian-free force-gradient integrators with $s \leq 6$ stages has been investigated in \cite{schäfers2024hessianfree} by deriving the minimum-error variants based on the efficiency measure \eqref{eq:EffErr} with appropriate weights for the additional error terms. In Table~\ref{tab:Hessian-free_overview}, the classification for the respective variants has been augmented by the stability threshold $z_*$ and the relative stability threshold $\EffStab$.

\renewcommand{\arraystretch}{1.2}
\begin{table*}[hp!]
    \centering
    \begin{tabular}{@{}c | l c c l r c c c c@{}}
        \toprule 
        & Integrator & $n_f$ & $n_g$ & $\mathrm{Err}_{\convergenceorder+1}$ & $\EffErr{\convergenceorder}$ & $z_*$ & $\EffStab$ & Remarks & Eqs. \\
        \midrule  
        \multirow{2}{*}{\rotatebox{90}{$\convergenceorder = 2$}} 
        & CAC & 1 & 1 & 0.0134 & 1.33 & 1.7791 & 0.5930 & \cite{omelyan2003symplectic} & \\
        & ACA & 1 & 1 & 0.00648 & 2.67 & 2.4495 & 0.8165 & \cite{omelyan2003symplectic} & \\
        \midrule
        \multirow{24}{*}{\rotatebox{90}{$\convergenceorder = 4$}} 
        & CABAC & 2 & 1 & 0.00334 & 1.17 & 2.4495 & 0.6124 & \cite{omelyan2003symplectic} & \\
        & BACAB & 2 & 1 & 0.000713 & 5.47  & 3.4641 & 0.8660 & \cite{suzuki1995new,chin1997symplectic} & \eqref{eq:BACAB} \\
        & CACAC & 2 & 2 & 0.000595 & 1.30  & 3.2714 & 0.5452 & \cite{omelyan2003symplectic} & \\
        & ACACA & 2 & 2 & 0.000715 & 1.08  & 2.9269 & 0.4878 & \cite{suzuki1995new,chin1997symplectic} & \\
        & CABABAC & 3 & 1 & 0.000855 & 1.87  & 3.0089 & 0.6018 & \cite{omelyan2003symplectic} & \\
        & ABACABA & 3 & 1 & 0.000141 & 11.31  & 3.1393 & 0.6279 & \cite{omelyan2003symplectic} & \eqref{eq:ABACABA} \\
        & BACACAB & 3 & 2 & 0.0000443 & 9.40  & 3.1259 & 0.4466 & \cite{omelyan2003symplectic} & \\
        & ACABACA & 3 & 2 & 0.0000823 & 5.06  & 3.0744 & 0.4392 & \cite{omelyan2003symplectic} & \\
        & CACACAC & 3 & 3 & 0.0000167 & 9.11  & 3.1079 & 0.3453 & \cite{omelyan2003symplectic} & \\
        & ACACACA & 3 & 3 & 0.0000123 & 12.42 & 3.1167 & 0.3463 & \cite{omelyan2003symplectic} & \\
        & BABACABAB & 4 & 1 & 0.0000634 & 12.17 & 3.1131 & 0.5189 & \cite{omelyan2003symplectic} & \\
        & CABABABAC & 4 & 1 & 0.000294 & 2.63 & 3.0233 & 0.5039 & \cite{omelyan2003symplectic} & \\
        & CABACABAC & 4 & 2 & 0.00000368 & 66.34 & 3.0883 & 0.3860 & \cite{omelyan2003symplectic} & \eqref{eq:CABACABAC} \\
        & BACABACAB & 4 & 2 & 0.00000649 & 37.60 & 3.1457 & 0.3932 & \cite{omelyan2003symplectic} & \eqref{eq:BACABACAB} \\
        & ABACACABA & 4 & 2 & 0.0000323 & 7.56 & 3.1366 & 0.3921 & \cite{omelyan2003symplectic} & \\
        & ACABABACA & 4 & 2 & 0.0000464 & 5.26 & 3.1036 & 0.3880 & \cite{omelyan2003symplectic} & \\
        & CACABACAC & 4 & 3 & 0.00000605 & 16.53 & 3.0630 & 0.3063 & \cite{omelyan2003symplectic} & \\
        & ACACACACA & 4 & 4 & 0.00000312 & 15.48 & 3.1277 & 0.2606 & \cite{omelyan2003symplectic} & \\
        & CABABABABAC & 5 & 1 & 0.0000165 & 25.22 & 3.1022 & 0.4432 & \cite{omelyan2003symplectic} & \\
        & ABABACABABA & 5 & 1 & 0.0000121 & 34.54 & 3.1110 & 0.4444 & \cite{omelyan2003symplectic} & \\
        & BACABABACAB & 5 & 2 & 0.00000320 & 47.60 & 3.1356 & 0.3484 & \cite{omelyan2003symplectic} & \\
        & BABACACABAB & 5 & 2 & 0.0000132 & 11.58 & 3.1213 & 0.3468 & \cite{omelyan2003symplectic} & \\
        & ABACABACABA & 5 & 2 & 0.00000127 & 120.03 & 3.1223 & 0.3469 & \cite{omelyan2003symplectic} & \eqref{eq:ABACABACABA} \\
        & ACABABABACA & 5 & 2 & 0.0000117 & 13.07 & 3.1127 & 0.3459 & \cite{omelyan2003symplectic} & \\
        \midrule
        \multirow{9}{*}{\rotatebox{90}{$\convergenceorder = 6$}} 
        & BACACACAB & 4 & 3 & 0.00150 & 0.000668 & 2.1844 & 0.2184 & \cite{omelyan2003symplectic} & \\
        & CABACACABAC & 5 & 3 & 0.0000147 & 0.0383 & 3.0957 & 0.2814 & \cite{omelyan2003symplectic} & \\
        & CACABABACAC & 5 & 3 & 0.0000264 & 0.0214 & 2.9559 & 0.2687 & \cite{omelyan2003symplectic} & \\
        & ACABACABACA & 5 & 3 & 0.00000607 & 0.0931 & 3.0426 & 0.2766 & \cite{omelyan2003symplectic} & \\
        & ABACACACABA & 5 & 3 & 0.000146 & 0.00388 & 3.2927 & 0.2993 & \cite{omelyan2003symplectic} & \\
        & BACACACACAB & 5 & 4 & 0.00000366 & 0.0566 & 3.1389 & 0.2415 & \cite{omelyan2003symplectic} & \\
        & ACACABACACA & 5 & 4 & 0.0000139 & 0.0149 & 3.0253 & 0.2327 & \cite{omelyan2003symplectic} & \\
        & CACACACACAC & 5 & 5 & 0.00000249 & 0.0353 & 3.1384 & 0.2092 & \cite{omelyan2003symplectic} & \\
        & ACACACACACA & 5 & 5 & 0.00000299 & 0.0294 & 3.1725 & 0.2115 & \cite{omelyan2003symplectic} & \\
        \bottomrule
    \end{tabular}
    \caption{Collection of conventional force-gradient integrators in their respective minimum-error variants. For all integrators, the number of force evaluations $n_f$ and force-gradient evaluations $n_g$ per time step, the leading error term $\mathrm{Err}_{\convergenceorder + 1}$, the efficiency measure $\EffErr{\convergenceorder}$, the stability threshold $z_*$, the relative stability threshold $\EffStab$, as well as a remark on the earlier appearance of the integrations, are stated. If available, the final column refers to the equations where the respective variant is defined.}
    \label{tab:force-gradient_overview}
\end{table*}

\begin{table*}[hp!]
    \centering
    \begin{tabular}{@{}c | l c c c c c c c c@{}}
        \toprule 
        & Integrator & $n_f$ & $n_g$ & $\mathrm{Err}_{\convergenceorder+1}$ & $\EffErr{\convergenceorder}$ & $z_*$ & $\EffStab$ & Remarks & Eqs. \\
        \midrule  
        \multirow{2}{*}{\rotatebox{90}{$\convergenceorder = 2$}} 
        & DAD & 1 & 1 & 0.0833 & 3.00 & 1.7791 & 0.8895 & \cite{omelyan2003symplectic,Hairer_McLachlan_Skeel_2009} & \\
        & ADA & 1 & 1 & 0.0417 & 6.00 & 2.4495 & 1.2247 & \cite{omelyan2003symplectic} & \\
        \midrule
        \multirow{28}{*}{\rotatebox{90}{$\convergenceorder = 4$}} 
        & DABAD & 2 & 1 & 0.00335 & 3.68 & 2.4495 & 0.8165 & \cite{omelyan2003symplectic} & \\
        & BADAB & 2 & 1 & 0.000728 & 16.96 & 3.4641 & 1.1547 & \cite{suzuki1995new,chin1997symplectic,yin2011improving} & \eqref{eq:BADAB} \\
        & ADADA & 2 & 2 & 0.000718 & 5.44 & 2.9269 & 0.7317 & \cite{suzuki1995new,chin1997symplectic} & \\
        & DADAD & 2 & 2 & 0.000625 & 6.25 & 3.2821 & 0.8205 & \cite{schäfers2024hessianfree} & \\
        & DABABAD & 3 & 1 & 0.000891 & 4.38 & 2.9829 & 0.7457 & \cite{schäfers2024hessianfree} & \\
        & ABADABA & 3 & 1 & 0.000149 & 26.19 & 3.1377 & 0.7844 & \cite{schäfers2024hessianfree} & \eqref{eq:ABADABA}, \eqref{eq:ABADABA_min-error} \\
        & ADABADA & 3 & 2 & 0.0000844 & 18.95 & 3.0745 & 0.6149 & \cite{schäfers2024hessianfree} & \\
        & BADADAB & 3 & 2 & 0.0000498 & 32.12 & 3.1261 & 0.6252 & \cite{schäfers2024hessianfree} & \\
        & DADADAD & 3 & 3 & 0.0000275 & 28.09 & 3.1082 & 0.5180 & \cite{schäfers2024hessianfree} & \\
        & ADADADA & 3 & 3 & 0.0000200 & 38.57 & 3.1157 & 0.5193 & \cite{schäfers2024hessianfree} & \\
        & DABABABAD & 4 & 1 & 0.000336 & 4.76 & 3.0907 & 0.6181 & \cite{schäfers2024hessianfree} & \\
        & BABADABAB & 4 & 1 & 0.0000651 & 24.57 & 3.1123 & 0.6225 & \cite{schäfers2024hessianfree} & \\
        & ADABABADA & 4 & 2 & 0.0000471 & 16.39 & 3.1039 & 0.5173 & \cite{schäfers2024hessianfree} & \\
        & ABADADABA & 4 & 2 & 0.0000346 & 22.32 & 3.1380 & 0.5230 & \cite{schäfers2024hessianfree} & \\
        & DABADABAD & 4 & 2 & 0.0000130 & 59.33 & 3.0851 & 0.5142 & \cite{schäfers2024hessianfree} & \\
        & BADABADAB & 4 & 2 & 0.0000105 & 73.45 & 3.1457 & 0.5243 & \cite{schäfers2024hessianfree} & \eqref{eq:BADABADAB}, \eqref{eq:BADABADAB_min-error} \\
        & DADABADAD & 4 & 3 & 0.0000101 & 41.06 & 3.1340 & 0.4477 & \cite{schäfers2024hessianfree} & \\
        & ADADADADA & 4 & 4 & 0.00000501 & 48.71 & 3.1265 & 0.3908 & \cite{schäfers2024hessianfree} & \\
        & DABABABABAD & 5 & 1 & 0.0000166 & 46.47 & 3.0999 & 0.5166 & \cite{schäfers2024hessianfree} & \\
        & ABABADABABA & 5 & 1 & 0.0000154 & 50.09 & 3.1050 & 0.5175 & \cite{schäfers2024hessianfree} & \\
        & BABADADABAB & 5 & 2 & 0.0000189 & 21.98 & 3.1289 & 0.4470 & \cite{schäfers2024hessianfree} & \\
        & ADABABABADA & 5 & 2 & 0.0000128 & 32.64 & 3.1130 & 0.4447 & \cite{schäfers2024hessianfree} & \\
        & BADABABADAB & 5 & 2 & 0.00000520 & 80.13 & 3.1371 & 0.4482 & \cite{schäfers2024hessianfree} & \eqref{eq:BADABABADAB} \\
        & ABADABADABA & 5 & 2 & 0.00000445 & 93.60 & 3.1239 & 0.4463 & \cite{schäfers2024hessianfree} & \eqref{eq:ABADABADABA} \\
        & DADABABADAD & 5 & 3 & 0.00000519 & 47.08 & 3.1355 & 0.3919 & \cite{schäfers2024hessianfree} & \\
        & DABADADABAD & 5 & 3 & 0.00000355 & 68.84 & 3.1381 & 0.3923 & \cite{schäfers2024hessianfree} & \\
        & ADABADABADA & 5 & 3 & 0.00000318 & 76.79 & 3.1358 & 0.3920 & \cite{schäfers2024hessianfree} & \\
        & ADADABADADA & 5 & 4 & 0.00000235 & 64.99 & 3.1283 & 0.3476 & \cite{schäfers2024hessianfree} & \\
        \midrule
        \multirow{3}{*}{\rotatebox{90}{$\convergenceorder = 6$}} 
        & BADADADAB & 4 & 3 & 0.00154 & 0.0055 & 2.1844 & 0.3121 & \cite{omelyan2003symplectic} & \\
        & BADADADADAB & 5 & 4 & 0.00000699 & 0.27 & 3.0727 & 0.3414 & \cite{schäfers2024hessianfree} & \\
        & ADADADADADA & 5 & 5 & 0.00000603 & 0.17 & 3.0806 & 0.3081 & \cite{schäfers2024hessianfree} & \\
        \bottomrule
    \end{tabular}
    \caption{Collection of Hessian-free force-gradient integrators in their respective minimum-error variants. For all integrators, the number of force evaluations $n_f$ and force-gradient evaluations $n_g$ per time step, the leading error term $\mathrm{Err}_{\convergenceorder + 1}$, the efficiency measure $\EffErr{\convergenceorder}$, the stability threshold $z_*$, the relative stability threshold $\EffStab$, as well as a remark on the earlier appearance of the integrations, are stated. If available, the final column refers to the equations where the respective variant is defined.}
    \label{tab:Hessian-free_overview}
\end{table*}
\FloatBarrier 
\onecolumn
\twocolumn
\bibliographystyle{elsarticle-num} 
\bibliography{refs}

\end{document}